\documentclass[twocolumn,showpacs,preprintnumbers,amsmath,amssymb,superscriptaddress,nofootinbib,prd]{revtex4}
\pdfoutput=1
\usepackage{amssymb}
\usepackage{mathrsfs}
\usepackage{graphicx}% input figure files
\usepackage{dcolumn}% Align table columns on decimal point
\usepackage{bm}% bold math
\usepackage{mathrsfs}

\begin{document}

\preprint{LAPTH-006/12}

\title{Probing neutralino dark matter in the MSSM \& the NMSSM with directional detection}

\author{D. Albornoz V\'asquez}
\affiliation{LAPTH, U. de Savoie, CNRS, BP 110, 74941 Annecy-Le-Vieux, France}
\affiliation{CNRS, UMR7095, Institut d'Astrophysique de Paris, F-75014, Paris, France}
\author{G. B\'elanger}
\affiliation{LAPTH, U. de Savoie, CNRS, BP 110, 74941 Annecy-Le-Vieux, France}
\author{J. Billard}
\author{F. Mayet}
\affiliation{Laboratoire de Physique Subatomique et de Cosmologie, Universit\'e Joseph Fourier Grenoble 1,
  CNRS/IN2P3, Institut Polytechnique de Grenoble, Grenoble, France}

\date{\today}

\begin{abstract}
We investigate the capability of directional detectors to probe neutralino dark matter in the Minimal Supersymmetric Standard Model and the Next-to-Minimal Supersymmetric Standard Model with parameters defined at the weak scale. We show that directional detectors such as the future MIMAC detector will probe spin dependent dark matter scattering on nucleons that are beyond the reach of current spin independent detectors. The complementarity between indirect searches, in particular using gamma rays from dwarf spheroidal galaxies, spin dependent and spin independent direct search techniques is emphasized. We comment on the impact of the negative results on squark searches at the LHC. Finally, we investigate how the fundamental parameters of the models can be constrained in the event of a dark matter signal.   
\end{abstract}

\pacs{95.35.+d, 14.80.Ly}

\maketitle

\section{Introduction}
The existence of cold Dark Matter (DM) is inferred from a large number of astrophysical observations at various scales.
Locally, the large discrepancy between Newton's law of gravitation and the observed rotation curves indicates that  
spiral galaxies should be embedded in a DM halo~\cite{halo,halo.mw}.\\
Candidates for this class of to-be-discovered particles naturally arise from extensions of the standard model of particle
physics ({\it e.g} Supersymmetry),  as long as  an {\it ad-hoc} discrete symmetry is invoked to preserve the proton 
stability. Indeed, the lightest particle of the secluded sector (in Supersymmetry: the lightest supersymmetric particle (LSP)) becomes stable and can  be a good 
candidate for the cold DM present in the Universe especially if it is a weakly interacting massive particle (WIMP). Numerous DM candidates have been 
proposed in extensions of the standard model and include fermions, scalars or gauge bosons~\cite{Bertone:2010zz}.\\
Tremendous experimental efforts on a host of techniques have been made in the field of WIMP  detection, the main experimental issue  being the WIMP-background discrimination.  Directional detection of galactic DM has been   proposed  as  a powerful tool to identify genuine WIMP events as such~\cite{spergel}.
Recent studies have shown that, within the framework of 
dedicated statistical data analysis, a low exposure directional detector could lead either to   a high significance discovery 
of galactic DM~\cite{billard.disco,billard.profile,green.disco} or to a conclusive exclusion~\cite{billard.exclusion}, 
depending on the value of the unknown  WIMP-nucleon cross section. In the case of a high significance detection, it would  also be possible to go further 
and to constrain   the  WIMP properties, both from particle physics (mass, cross section) 
and  galactic halo (velocity dispersions), within the framework of a model independent analysis~\cite{billard.ident}.
To achieve this would require a rather high spin dependent (SD) cross section (of the order of $10^{-4} \rm pb$) and a low WIMP mass, comparable to the target 
nucleus mass, as a matter of fact below the electroweak scale. 
Most projects of directional detection~\cite{white} are low pressure Time Projection Chambers (50-100 mbar). It follows that there is an
%jb
obvious size limitation, which in turn  implies that  directional detectors cannot scale up to ton-scale experiment as most of the current direct detection 
projects~\cite{Ahmed:2011gh,Aprile:2011hi,Aalseth:2011wp,Angloher:2011uu,cdms-spin,simple,edelweiss-spin,coupp,kims,naiad,picasso,xenon10,zeplin}. Then, to be
competitive and complementary with planned and existing direct detectors, directional detection should focus on SD interaction for which a large fraction of
the parameter space could be probed by
planned directional experiments (see Sec.\ref{sec:directional}).\\ 
%\cite{FERMILAB-CONF-09-801-E, arXiv:1107.2155}.\\
DM models which can best be probed by directional SD detection are
the ones with Majorana particles such as the neutralino in the Minimal Supersymmetric Standard Model (MSSM) or its extensions. 
Not only can the neutralino be light (below $100$ GeV) but also its Majorana nature implies that  different processes contribute to the SD or spin independent (SI) interaction, namely the Z exchange contributes only to the SD cross section. Both types of detection modes are therefore complementary.
Vector DM candidates can also lead to a signal in the SD mode which is not directly correlated with the SI mode.  
However  in most models studied previously (for example UED or little Higgs~\cite{hep-ph/0602206,hep-ph/0603077}) the DM tends to be heavier than 
$100$ GeV  (even heavier 
than $1$ TeV in the minimal UED case~\cite{Belanger:2010yx}) making it difficult to extract the DM properties for direct detection.
Models with Dirac fermion DM get a contribution from Z exchange for both SI and
SD, these DM candidates are therefore best probed by SI where one can take advantage of the coherence effect.\\ 
We will therefore consider only models with a Majorana DM particle, more specifically the neutralino in 
the MSSM and in one of its extensions, the Next-to-Minimal Supersymmetric Standard Model (NMSSM)~\cite{Ellwanger:2009dp}.  
The parameter space of these models consistent with collider physics, precision measurements, DM relic density as well as 
DM direct 
%gb and indirect 
searches were examined in several studies assuming  either some GUT scale relations among the 
fundamental parameters~\cite{cmssm,arXiv:0806.1923,arXiv:1106.2529,arXiv:1011.6118,arXiv:1111.6098,arXiv:0906.4911,arXiv:0903.0555,arXiv:0910.3950,nonuni}  or the more general case of parameters defined at the electroweak scale~\cite{arXiv:0802.3384,arXiv:0812.0980,arXiv:0904.2548,arXiv:0906.5048,arXiv:1109.5119,arXiv:1110.3726}. 
%gb
The potential of indirect detection for probing these models was examined for example 
in~\cite{hep-ph/0401186,hep-ph/0507086,hep-ph/0607266,arXiv:0707.0622,arXiv:0810.5292,arXiv:0909.3300,arXiv:1106.0768} 
and the complementarity between detection techniques were investigated in~\cite{hep-ph/0405210,arXiv:0710.0553,arXiv:1011.4514,arXiv:1111.2607}.
In particular, in a previous study~\cite{Vasquez:2010ru,Vasquez:2011yq} the allowed parameter space of the MSSM and of the NMSSM 
with light  neutralinos (up to $50$ GeV)  was explored using a MCMC approach. Furthermore indirect detection of DM 
through the flux of photons from dwarf spheroidal galaxies, as well as direct detection limits 
from XENON100 were shown to be complementary and to further constrain the parameter space.\\  
In this study we first determine the parameter space of the MSSM and NMSSM  that can be explored by directional SD
interactions with the future MIMAC detector~\cite{mimac}.  For this we consider in both models 
the whole spectrum of masses for neutralinos although we use  a restricted set of free parameters at the electroweak scale.
We then compare this with limits from SI and indirect detection searches, namely gamma-rays.
Finally, we
investigate how the fundamental parameters of the models can be constrained in the event of a DM signal. 

The paper is organized as follows. In section~\ref{sec:directional} we discuss the potential of SD  directional detector in probing DM.
In section~\ref{sec:xs} we discuss the expectation for the SD cross section.
The parameter space of the model considered and the scanning method used are presented in Section~\ref{sec:susyconf}.
Our results for the  SI/SD direct searches as well as for the flux of photons relevant for  indirect 
searches  are presented in Section~\ref{sec:results} highlighting the complementarity between different techniques. 
The impact of directional detectors on probing the parameter space of the model is also addressed while
the impact of the LHC on these results is sketched in section~\ref{sec:lhc}. An investigation on how the observation of a signal 
could be used to determine some parameters of the supersymmetric model is carried in section~\ref{sec:disco}.\\ 

%fm
%Section VIII contains our conclusions.\\

\section{Directional detection framework}
\label{sec:directional}
\subsection{Directional detectors}
There is a worldwide effort toward the development of a large TPC (Time Projection Chamber) devoted to 
directional detection~\cite{white} and all current projects~\cite{mimac,dmtpc,drift,d3,emulsions,newage} 
face common experimental challenges and share a unique goal:  the simultaneous measurement of the energy ($E_r$) and the direction of the 3D track ($\Omega_r$) of low energy recoils, 
thus  allowing to evaluate the double-differential spectrum $\mathrm{d}^2R/\mathrm{d}E_r\mathrm{d}\Omega_r$ down to the energy threshold.
It is worth emphasizing that it is the lowest energy at which both the track and the energy can be retrieved which is the key
experimental issue for directional detection.  
It follows that, to maximize the track length,  the pressure of the gaseous detector 
must be as low as possible, leading to rather small detector masses as the volume cannot be arbitrarly large.
One may then come to the
conclusion that the directional detection strategy should focus on SD  interactions to be competitive with planned and existing
direct detectors~\cite{Ahmed:2011gh,Aprile:2011hi,Aalseth:2011wp,Angloher:2011uu,cdms-spin,
simple,edelweiss-spin,coupp,kims,naiad,picasso,xenon10,zeplin}.\\
Then, the ideal directional target is a nucleus with non-vanishing spin. Leading candidates include: $\rm ^{1}H$, $\rm ^{3}He$ and 
$\rm ^{19}F$  which has been early suggested as
a golden target for SD  DM searches~\cite{ellis.spin}.  $\rm CF_4$ is indeed planned  as a sensitive medium for 
most upcoming directional detectors~\cite{white}.\\  
In the following, as a working example,  we present the case for a low exposure (30 kg.year)  $\rm CF_4$ TPC, operated at 
low pressure and  allowing 3D track reconstruction, with sense recognition down to 5 keV. 
Such performance is taken as the ultimate limit for a directional detector. A complete overview of the effect of the main experimental issues of directional detectors, such as background contamination, energy threshold, 
sense recognition efficiency, angular and energy resolution on exclusion limits and discovery potential, are presented in~\cite{billard.profile,billard.exclusion}.

\subsection{Directional detection}
\subsubsection{Directional event rate and astrophysical inputs}
Directional detection depends crucially on the local WIMP velocity distribution~\cite{alenazi.directionnelle,Green:2010gw,Serpico:2010ae}. The isothermal sphere halo model is 
often considered but it is worth going beyond this standard paradigm when trying to account for all astrophysical uncertainties. 
The multivariate Gaussian WIMP velocity distribution corresponds to the generalization of the standard isothermal sphere with a density profile 
$\rho(r)\propto 1/r^2$, leading to a smooth WIMP velocity distribution,  
a flat rotation curve and no substructure. The WIMP velocity distribution in the laboratory frame is   given by,
\begin{equation}
f(\vec{v}) = \frac{1}{(8\pi^3\det{\boldsymbol\sigma}^2_v)^{1/2}}\exp{\left[-\frac{1}{2}(\vec{v} - \vec{v}_{\odot})^T {\boldsymbol\sigma}^{-2}_v(\vec{v} - \vec{v}_{\odot})\right]}
\end{equation}
where ${\boldsymbol\sigma}_v = \text{diag}[\sigma_{x}, \sigma_{y}, \sigma_{z}]$ is the velocity dispersion tensor 
assumed to be diagonal in the Galactic rest frame ($\hat{x}$, $\hat{y}$, $\hat{z}$) and $\vec{v}_{\odot}$ is the Sun's velocity vector with respect to
the Galactic rest frame. When neglecting the Sun peculiar velocity and the Earth orbital 
velocity about the Sun,  $\vec{v}_{\odot}$ corresponds to the detector velocity in
the Galactic rest frame and is taken to be $v_{\odot} = 220$ km.s$^{-1}$ along the $\hat{y}$ axis pointing toward the constellation Cygnus at 
($\ell_{\odot} = 90^{\circ}$, $b_{\odot} = 0^{\circ}$), where $\ell$ and $b$ are the Galactic latitude and longitude. This way, we can consider the three velocity dispersions
along the three axis as nuisance parameters in order to take into account the effect of anisotropic WIMP velocity distribution when deriving the discovery potential of upcoming
directional detectors. The anisotropy is defined by the $\beta$ parameter as:
\begin{equation}
\beta = 1 - \frac{\sigma^2_y + \sigma^2_z}{2\sigma^2_x}
\end{equation}
Hence, $\beta = 0$ corresponds to an isotropic WIMP velocity distribution while $\beta <0$ corresponds to a tangential anisotropy and $\beta >0$ to a radial anisotropy. 
Using the
parametrization of the three velocity dispersions used in this study (see table~\ref{tab:prior}),
 we are considering WIMP velocity distributions with $\beta = 0 \pm 0.25$ (68\% C.L.). This range is
compatible with recent results from N-Body simulations which found $\beta = 0 - 0.4$~\cite{nezri} and observations 
which found a non vanishing $\beta$ parameter in the solar neighbourhood~\cite{Hansen:2004qs,Smith:2009kr}.\\

The directional recoil  rate  is given by~\cite{gondolo}:
\begin{equation}
\frac{\mathrm{d}^2R}{\mathrm{d}E_r\mathrm{d}\Omega_r} = \frac{\rho_0\sigma_0}{4\pi m_{\chi}m^2_r}F^2(E_r)\hat{f}(v_{\text{min}},\hat{q}),
\label{directionalrate}
\end{equation}
with $m_{\chi}$ the WIMP mass, $m_r$ the WIMP-nucleus reduced mass, $\rho_0$   the local DM density, $\sigma_0$   the
WIMP-nucleus elastic scattering cross section, $F(E_r)$  the form factor  (using the axial expression from~\cite{lewin}),  
$v_{\text{min}}$ the   minimal WIMP velocity required to produce a
nuclear recoil of energy $E_r$ and $\hat{q}$ the direction of the recoil momentum. 
Finally, $\hat{f}(v_{\text{min}},\hat{q})$ is the three-dimensional Radon transform of the WIMP 
velocity distribution $f(\vec{v})$, see~\cite{gondolo} for more details.\\
As one can see from eq.~\ref{directionalrate}, the directional rate is directly proportional to the local DM density at Solar radius ($\rho_0$) which is also subject to
important uncertainties. Following~\cite{billard.profile} we consider $\rho_0$ as a nuisance parameter, using $\rho_0 = 0.3 \pm 0.1$ GeV/cm$^3$.
 We used a mean value of 0.3 GeV/cm$^3$ for the sake of comparison between the different direct searches experiments.\\ The last astrophysical uncertainties to be 
 considered when deriving the discovery potential of upcoming directional detectors is the the velocity of
the Solar system orbit's in the Galactic rest frame taken as $v_{\odot} = 220 \pm 30$ km/s.\\

It is worth noticing that other astrophysical uncertainties like the escape velocity could be taken into account when assessing the sensitivity of a given direct detection
experiment. However, 
 the escape velocity is not considered in this study as we are
dealing with a low mass target material ($^{19}$F) associated to a low energy threshold. Indeed, as the minimal speed required to produce a 5 keV recoil energy is about 
$v_{\rm min} = 130$ km/s for a WIMP mass of 100 GeV/c$^2$, while the mean WIMP velocity in an earth based detector is about 300 km/s, we can deduce that
 the effect of a finite escape velocity will be negligible.\\

As a conclusion, accounting for uncertainties on the 
 astrophysical parameters is a step beyond 
the ``standard DM model", {\it i.e.} isotropic isothermal DM halo, with 
fixed values of density and Sun's circular velocity.  Evaluating the properties of  the DM halo is indeed still 
a subject of debates. The numerical values and uncertainties corresponding to these astrophysical inputs are given in table~\ref{tab:prior} (see~\cite{billard.profile} for a
detailed discussion).

\setlength{\tabcolsep}{0.1cm}
\renewcommand{\arraystretch}{1.4}
\begin{table}[t]
\begin{center}
\hspace*{-0.5cm}
\begin{tabular}{|c||c|}
\hline
 Nuisance parameters  &   Gaussian parametrization \\ \hline \hline  
 $\rho_0 \ {\rm [GeV/c^2/cm^{3}]}$  &  $0.3\pm0.1$\\ \hline  
 $v_{\odot} \ {\rm [km/s]}$ & $220 \pm 30$ \\ \hline 
 $\sigma_x \ {\rm [km/s]}$ & $220/\sqrt{2} \pm 20$  \\ \hline 
 $\sigma_y \ {\rm [km/s]}$ & $220/\sqrt{2} \pm 20$  \\ \hline  
 $\sigma_z \ {\rm [km/s]}$ & $220/\sqrt{2} \pm 20$  \\ \hline 
\end{tabular}
\caption{Gaussian parametrization (mean and standard deviation) of the different astrophysical nuisance parameters.}
\label{tab:prior}
\end{center}
\end{table}
\renewcommand{\arraystretch}{1.1}

\begin{figure}[htb]
\includegraphics[scale=0.47,natwidth=6cm,natheight=6cm]{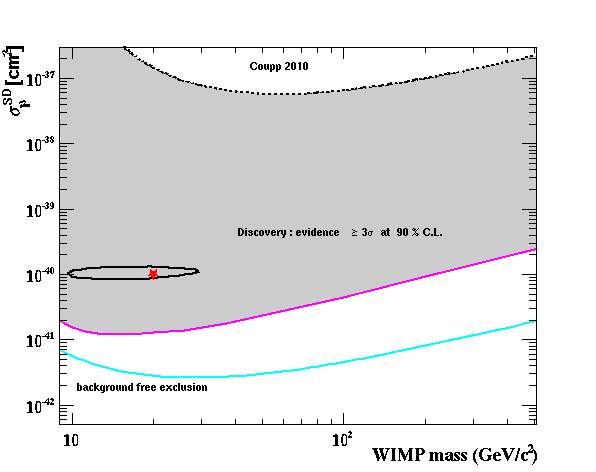}
\caption{
Spin dependent cross section on proton (cm$^2$) as a function of the WIMP mass ($\rm GeV$)
in the case of the pure-proton approximation~\cite{tovey} showing
the sensitivity of a forthcoming 30 kg.year  directional experiment (solid red line). It is defined 
as the minimal cross-section required to obtain a DM discovery with a significance
greater than the 3$\sigma$ level at 90\% C.L.~\cite{billard.profile}. 
For the same exposure, the dotted black line presents the projected exlusion limit~\cite{billard.exclusion} in the background free case. 
  The red star is the input value of the benchmark model
and the black contour is the 68\%  contour level obtained with the MCMC analysis~\cite{billard.ident}.
The exclusion limit from COUPP-2010~\cite{coupp} (black dotted line) is also presented.
}  
\label{fig:prospect1}
\end{figure}

\subsubsection{Dark Matter properties inferred from directional detection}
\label{sec:DMprop}
As highlighted by several recent studies~\cite{billard.disco,green.disco,billard.exclusion,billard.ident}, directional detection 
may be a powerful tool to discriminate between the DM signal and the
background one. Indeed, the correlation between the main incoming direction of the recoiling events in galactic coordinates with the Solar system orbit's around
the Galactic Center has been shown to be a strong and convincing proof in favor of a DM detection. This kind of discrimination is even more relevant when
considering the recent results from DAMA, CoGeNT and CRESST experiments which have observed candidate events which origin are difficult to assess without directional
information.\\
At first, one may think of using directional detection to set exclusion limits. Several methods have been proposed~\cite{henderson,billard.exclusion}. Figure~\ref{fig:prospect1} presents the expected rejection limit for a 30 kg.year
CF$_4$ directional detector, in the background-free case~\cite{billard.exclusion} for which a standard limit has been derived using the classical Poisson statistics
associated to a 0 observed event.
It allows to reach $\sim 10^{-6} \ {\rm pb}$, noticing
that for highly background-contaminated data ($\sim 10 \ {\rm kg^{-1}.year^{-1}}$) the result would be about an order of magnitude higher.\\
However, directional detection may be used to go beyond the standard exclusion limit strategy. Indeed, it may allow 
to discover DM~\cite{billard.disco,green.disco},   the proof of discovery being the fact that 
the signal points to the direction of  the  constellation Cygnus (to which the solar system's velocity vector is pointing).
Hence, the goal is to identify a genuine WIMP signal as such.
Using a frequentist profile likelihood ratio test statistics, and taking into account astrophysical and experimental uncertainties, 
one can determine the sensitivity of a given directional experiment~\cite{billard.profile}. It is defined, in our case, 
as the minimal cross-section required to obtain a DM detection with a significance
greater than the 3$\sigma$ level ain 90\% of the experiments.
The expected sensitivity of a directional detector filled of CF$_4$  with a low exposure (30 kg.year) and a 
5 keV energy threshold  is displayed in fig.~\ref{fig:prospect1}. Such a directional detector should be able to 
reach a sensitivity down to a SD cross-section of 10$^{-5}$ pb for a WIMP mass of 20 GeV/c$^2$. As one can see from the pink and cyan curves in fig.~\ref{fig:prospect1}, the
distance between the background free exclusion limit and the discovery sensitivity is lower at low WIMP mass than at high WIMP mass. This effect, highlights
the fact that directional detection is more sensitive to low WIMP mass as fewer number of events are required to reach a high significance detection for light WIMPs than for
heavy WIMPs. This is of a major interest when considering the low WIMP mass issue as directional detection could bring valuable information to discriminate between a genuine
WIMP detection and an unexpected background contamination of the DAMA, CoGent and CRESST experiments.

For high WIMP-nucleon SD cross sections, it is also possible to go even further~\cite{billard.ident}.  
With the help of a high dimensional mutivariate analysis, it is possible  to identify a WIMP
with directional detection. It has been shown that dedicated analyses of 
simulated pseudo-data of a 30 kg.year CF$_4$ directional detector would allow to constrain  
the  WIMP properties, both from particle physics ($m_\chi, \sigma_p^{SD}$) and 
galactic halo (velocity dispersions). For instance, for a benchmark model 
($m_\chi= 20 \ {\rm GeV/c^2}, \sigma_p^{SD}=10^{-4} \ {\rm pb}$), the constraints would be the following
\begin{align*}
& m_{\chi}   =  19.9^{+2.7}_{-8.8} \ {\rm GeV/c^2} \ (68\% \ {\rm CL}),  \\
& \log_{10}(\sigma_p)  =   -3.97 \pm 0.06 \ (68\% \ {\rm CL}).    
\end{align*}
Figure~\ref{fig:prospect1} presents  the 68\%   contour level in the ($m_{\chi},\sigma_n$) plane. 
This is indeed a model-independent measurement --as the velocity dispersions are set as free parameters within the framework of a multivariate Gaussian velocity distribution-- of the WIMP properties, consistent
with the input values and with a rather small dispersion.\\
To assess the interest of directional detection, either discovery or exclusion, in the following we explore the MSSM and NMSSM
parameter space in order to check if some models, excluded neither by colliders nor cosmology, would lie in the regions of interest.

\section{Spin dependent elastic scattering interactions}
\label{sec:xs}
 
\subsection{Neutralino-nucleon spin dependent cross section}\label{sec:xs1}
\begin{figure}[bt]
\centering
\includegraphics[width=0.45\textwidth ,natwidth=6cm,natheight=6cm]{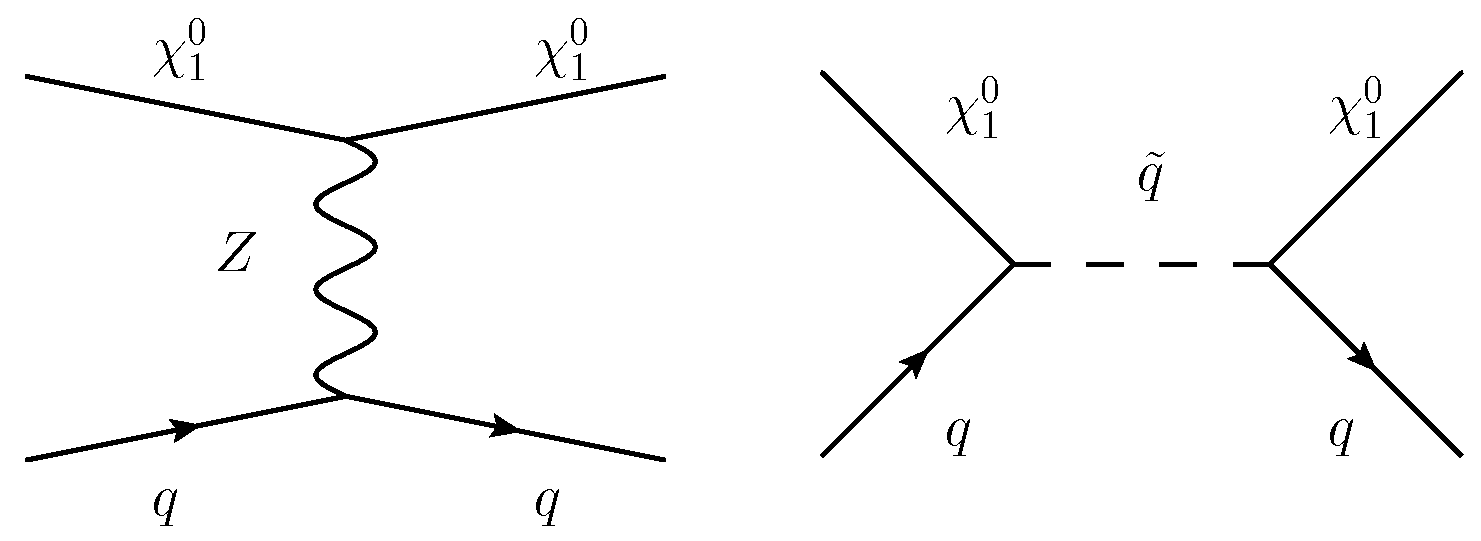}
\caption{Feynman diagrams contributing to spin dependent elastic scattering of neutralinos on nucleons.}
\label{fig:Feynman}
\end{figure}
In the non-relativistic approximation, the WIMP-nucleon interaction 
is composed of two contributions: the spin dependent (axial) and the spin independent one (scalar)~\cite{goodman}. 
Two diagrams contribute to the SD interactions of a neutralino: the Z exchange and the squark exchange 
(see Fig.~\ref{fig:Feynman}). 
The former relies on the Higgsino component of the LSP. When the Z contribution dominates, the amplitude for nucleon ($a_N$) is directly proportional to 
the neutralino Z coupling, 
\begin{equation}
a_N=-\left( \Delta_u^N-\Delta_d^N-\Delta_s^N \right)  \left( N_{13}^2- N_{14}^2 \right) \nonumber
\end{equation}
where $N_{13}, N_{14}$ are the higgsino-$d$ and higgsino-$u$ components of the LSP and the coefficients $\Delta_q^N$ describe the contribution of a quark $q$ to the spin 
of the nucleon. 
With
\begin{eqnarray}
\Delta_u^p&=&0.842\pm 0.012 \;\;\; \Delta_d^p=-0.427\pm 0.013\nonumber\\
\Delta_s^p&=&-0.085\pm 0.018 \nonumber
\end{eqnarray}
one can show~\cite{Belanger:2008gy} that the two amplitudes have opposite signs and that $a_p/a_n=-1.14\pm 0.03$ independently of the parameters of the model.
 If the squark exchange is important, the relation between amplitudes are shifted and can even have the same sign.
 Note that the relative sign of $a_p$ and $a_n$ is a key point for SD direct searches as it could lead to either constructive or destructive interference
 between the two amplitudes, depending on the spin content of the considered nucleus target (see sec.~\ref{sec:SDCross}).
 The squark contribution to the amplitude reads
\begin{eqnarray}
\frac{a_N}{8M_Z^2 c_W^2}&=&    \frac{ t^2_W {N_{11}}^2}{9}    \left( \frac{4\Delta_u^N}{m^2_{\tilde u_R}-m^2_{\tilde \chi}}+\frac{\Delta_d^N+\Delta_s^N}{{m^2_{\tilde d_R}-m^2_{\tilde \chi}}} \right)  \nonumber\\
&+& (N_{12}+\frac{t_W}{6} N_{11})^2\frac{\Delta_u^N}{4(m^2_{\tilde u_L}-m^2_{\tilde \chi})} \nonumber\\
&+&(N_{12}-\frac{t_W}{6} N_{11})^2\frac{\left(\Delta_d^N+\Delta_s^N\right)}{4(m^2_{\tilde d_L}-m^2_{\tilde \chi})} 
\end{eqnarray}
where $N_{11}$, $ N_{12}$ are respectively the bino and wino component of the LSP and $c_W=\cos\theta_W$. 
In the limit of  an almost pure bino LSP, which is usually the case for a light neutralino, the amplitude simplifies to
\begin{eqnarray}
\frac{a_N}{8M_Z^2 c_W^2}&=&  \frac{ t^2_W {N^2_{11}}}{9}\left[
\frac{\left(\Delta_u^N+\Delta_d^N+\Delta_s^N\right)}{4(m^2_{\tilde q_L}-m^2_{\tilde \chi})}\right.\nonumber\\
 &+&\left. \frac{4\Delta_u^N}{m^2_{\tilde u_R}-m^2_{\tilde \chi}} +\frac{\Delta_d^N+\Delta_s^N} {m^2_{\tilde d_R}-m^2_{\tilde \chi}}
\right]
 \end{eqnarray}
 If furthermore the  L and R squark masses are equal, the dominant contribution comes from $\tilde{u}_R$ and $\tilde{d}_R$ respectively as the neutralino/quark/squark coupling is
 proportional to the quark hypercharge which is larger for RH quarks. The ratio of proton and neutron amplitudes is then
$a_p/a_n= - 3.38\pm0.22$.  In the case where  $\tilde{q}_L$ gives the largest contribution the ratio of amplitudes is positive with   $a_p/a_n= 1$.
  The squark exchange is however usually suppressed as compared to the Z exchange because of the higher mass scale involved, $m_{\tilde q} \gg M_Z$. It can be dominant when
  the higgsino component is very small leading to a suppressed coupling of the neutralino to the Z and to small cross sections. In fact for the squark contribution to be
  relevant for the discovery reach of directional detection ($\sigma_p \simeq 5 \times 10^{-6}$ pb), one can estimate that the common squark mass has  to be below roughly 620 GeV when the LSP is a  light bino ($N_{11}=1$). Such squark masses are excluded by LHC if the dominant decay mode of the squark is into $q+LSP$~\cite{Aad:2011ib}.  Note that  a cancellation between the squark and Z exchange can lead to $\sigma_p\gg \sigma_n$ or $\sigma_p\ll \sigma_n$, we however stress that this occurs typically only for small cross sections if the squarks are near the TeV scale as will be discussed further in Section~\ref{sec:lhc}.

\subsection{Constraining the Spin Dependent cross section}
\label{sec:SDCross}
The Spin Dependent (SD) cross-section (at zero momentum transfer) on a nucleus $^{A}{\mathrm{X}}$ is given by
\begin{equation}
\sigma^{SD}(^{A}{\mathrm{X}}) = {32 \over \pi} G^2_F  \,\mu^2_A  {J+1 \over J} \left(a_p <S_p> +\; a_n <S_n> \right)^2 
\label{eq:xsnoyau}
\end{equation} 
where $G_F$ is the Fermi constant, $\mu^2_A$ is the WIMP-nucleus reduced mass, J the total nuclear spin
and $<S_{p,n}>\  = \ <N \mid S_{p,n} \mid N \ >$ the  neutron/proton spin content of the target  nucleus.  
To be sensitive to the SD interaction, one may either choose a pure non-zero spin nucleus target ($\rm ^{3}He, ^{19}F$) or rely on the isotopic
fraction of the chosen target. This way, scalar detection experiments may also impose constraints on 
SD interaction thanks to the small fraction of odd-A target nucleus in the sensistive medium 
({\it e.g.} $\rm ^{73}Ge$ for Ge-based detectors or $\rm ^{129,131}Xe$ for Xenon-based ones).

The result of a DM experiment, either a discovery or an exclusion, implies a constraint on 
$\sigma^{SD}(^{A}{\mathrm{X}})$ which must then be converted on a constraint on 
$\sigma_{p,n}$ for the sake of comparison between various DM experiments using different target nuclei.
This is however not straightforward as the amplitude on both proton and neutron are involved in (\ref{eq:xsnoyau}).\\  
%fm ok
%gb: "shall be" ne me plaisait ca laissait entendre que c'etait ce qu'on allait faire
%In the pure-nucleon coupling approximation   ($a_n=0$ or $a_p=0$), the  cross-section limit on the nucleus 
%  shall be converted into a cross-section limit  on proton (resp. neutron).
For simplicity, the pure-nucleon coupling approximation   ($a_n=0$ or $a_p=0$) is often used. 
This method is however WIMP-model dependent as there is no particular reason
for one coupling to vanish, indeed this does not occur in the MSSM.
A model-independent method  has been proposed by D.~R.~Tovey {\it et al.}~\cite{tovey}  to enable
comparison amongst SD direct searches of DM.  
For a given WIMP mass, an exclusion limit on 
$\sigma^{SD}(^{A}{\mathrm{X}})$ is then translated into a constraint on $\sigma_p$ and  $\sigma_n$, as: 
$$
\Big(\sqrt{\sigma_p} \pm \frac{<S_n>}{<S_p>} \sqrt{\sigma_n} \Big)^2 < \sigma_p^{lim} 
$$
where $\sigma_p^{lim}$ is the limit on WIMP-proton cross section obtained in the pure proton approximation.\\
The limit is expressed, for a given WIMP mass, in the nucleon cross-section plane ($\sigma_p, \sigma_n$) and divided in two cases: 
``constructive" and ``destructive", whether the diffusion amplitude on proton and neutron add up coherently or not, taking into account   
the relative  sign of $<S_{p}>$ and $<S_{n}>$. Hence, the exclusion in the ($\sigma_p, \sigma_n, m_\chi$) space does not 
depend on a particular WIMP model.\\
As outlined in~\cite{Cannoni:2011iu}, setting a limit on the SD interaction requires to neglect the SI one, which is 
not always justified. In particular, in the case of Fluorine, the
SD rate can be dominant over the SI, but this has to
be checked in each particular WIMP model.

\subsubsection{Proton-based versus neutron-based detectors}
The knowledge of     the expectation values of the spin content of the proton and neutron 
 within the nucleus ($<S_{p,n}>\  = \ <N \mid S_{p,n} \mid N \ >$) is a key issue for SD detection of DM. 
 The WIMP couples  mainly to the spin of the unpaired proton ({\it e.g.} $\rm ^{19}F$) or to the one of the unpaired neutron 
({\it e.g.}  $\rm ^{3}He$), leading to a contraint  on either the proton or neutron diffusion amplitude  ($a_n, a_p$). 
One may then distinguish {\em proton-based} ($\rm ^{19}F$, $\rm ^{23}Na$, $\rm ^{27}Al$, $\rm ^{35}Cl$, $\rm ^{127}I$) and 
{\em neutron-based} ($\rm ^{3}He$, $\rm ^{73}Ge$, $\rm ^{129}Xe$) SD experiments which shall present an obvious complementarity~\cite{Moulin:2005sx,thesemanu,giuliani}.\\
However in practice, the spin of the target nucleus is carried both by constituent neutrons and
 protons and the comparison between SD experiments is not straightforward.\\
Detailed nuclear shell-model calculations have been developped and  the accuracy of the  $<S_p>$ and $<S_n>$ evaluation is 
assessed by comparing, within the same shell model,  the predictions on the magnetic
moment and the energy spectra for the nuclear lowest eigenstates with experimental values. We refer the reader to~\cite{bednyakov} for 
a comprehensive discussion on the subject.\\
Several collaborations have published exclusion limits on SD WIMP-nucleon cross-sections~\cite{cdms-spin,coupp,edelweiss-spin,kims,naiad,picasso,simple,xenon10,zeplin}. In the following, we use (see fig.~\ref{fig:prospect1}): 
\begin{itemize}
\item  for proton-based detector:   COUPP-2010~\cite{coupp}, obtained with a 2-liter  $CF_3I$ 
Bubble Chamber, with a 28.1 kg.day effective exposure, 
\item  for neutron-based detector: XENON10~\cite{xenon10}, obtained with 5.4 kg of fiducial
liquid xenon, with a 136 kg.day effective exposure.
\end{itemize}

\subsubsection{The case of $^{19}F$}
\begin{figure*}[t]
\hspace*{-1.1cm}
\includegraphics[scale=0.47,natwidth=6cm,natheight=6cm]{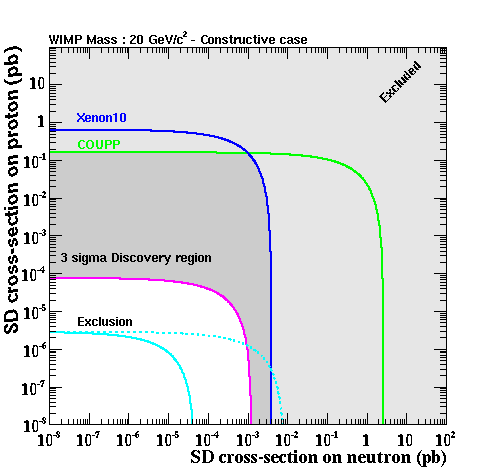}
\includegraphics[scale=0.47,natwidth=6cm,natheight=6cm]{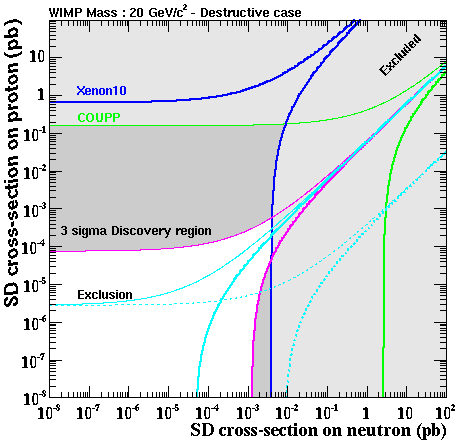}
\caption{Left (resp.
Right) panel present the constraints in the constructive (resp. destructive) case,
for a 20  ${\rm GeV}$ WIMP. The light grey area is the region already excluded by the most constraining experiments 
(COUPP-2010~\cite{coupp}  and XENON10~\cite{xenon10}). For a 30 kg.year $CF_4$ directional detector, 
the dark grey area is the $3 \sigma$  discovery region, while the solid light curve labelled ``exclusion" is the background-free projected
limit. Dashed light curve presents the same result for the alternative $^{19}F$ spin content values~\cite{divari}.}
\label{fig:sigsig-all}
\end{figure*}

In the following we emphasize that some discrepancies remain on the spin content of $^{19}F$. 
The value of the spin content of $^{19}F$ is predicted by two 
authors~\cite{pacheco,divari}, see tab.~\ref{tab:cont.spin}. It is dominated by the proton content but the
neutron one varies by more than one order of magnitude. Values from A.~F.~Pacheco \& D.~Strottman  are widely used. 
However, it is noteworthy  that the evaluation from P.~C.~Divari {\it et al.} is obtained by shell model calculations 
using the Wildenthal interaction~\cite{wildenthal}, which is known to reproduce accurately 
many nuclear observables.\\
Nonetheless, to allow a fair comparison with existing published results~\cite{coupp,picasso}, 
the values $<S_p>\ =0.441$ and  $<S_n>\ =-0.109$ from~\cite{pacheco} are chosen, noticing that the other 
choice would only mildly alter the result on the proton cross-section limit but would scale up   the result 
for neutron cross-section   by two orders of magnitude (see. fig.~\ref{fig:sigsig-all}). 
We leave the question open, highlighting the fact that results on  $^{19}F$ must be treated with caution owing to the 
nuclear-shell model dependence.\\
However, it is worth emphasizing that, in the case of  $^{19}F$, the spin contents having opposite sign, one expects a 
constructive interference between the proton and neutron amplitudes, as  they usually also have opposite sign (sec.~\ref{sec:xs1}).
 
\begin{table}[t]
\begin{center}
\begin{tabular}{cccccc}
\hline
\hline
Model &   $<S_p>\;$ & $<S_n>\;$ & Ref. \\ \hline \hline 
odd-group & 0.5 & 0. & \\
Pacheco \& Strottman & 0.441 & -0.109 &~\cite{pacheco} \\
Divari {\it et al.} & 0.475 & -0.0087 &~\cite{divari} \\ \hline 
\end{tabular}
\caption{Spin content of $^{19}F$ in various nuclear shell models. Henceforth, values from~\cite{pacheco} are 
used.}
\label{tab:cont.spin}
\end{center}
\end{table}

Figure~\ref{fig:sigsig-all} presents, for a 20  ${\rm GeV}$ WIMP,  the sensitivity of the considered directional detector.
The light grey area is the region already excluded by the most constraining experiments 
(COUPP-2010~\cite{coupp}  and XENON10~\cite{xenon10}). In the case of XENON10, the curve is derived from~\cite{xenon10} considering a pure
$^{129}Xe$ detector, which is justified as the contribution of odd xenon  isotopes are very close. In the case of COUPP, the curve is 
derived from~\cite{coupp} considering a pure
$^{19}F$ detector, {\it i.e.} neglecting the contribution of $^{127}I$ for such a small WIMP mass. The exclusion limit and the $3 \sigma$  discovery region, 
for a 30 kg.year $CF_4$ directional detector are presented on fig.~\ref{fig:sigsig-all}. It corresponds to a rather large region in the
parameter space, well below current limits.\\

\section{Testing supersymmetric configurations}
\label{sec:susyconf}
In order to establish the reach of directional detection for probing supersymmetric models, we take into account various constraints that have an impact on the parameter space.

\subsection{Parameter space}

We consider two models, both with parameters defined at the electroweak scale, the MSSM and the NMSSM. The free parameters we take in the MSSM are the same as in~\cite{Vasquez:2011yq}. We assume minimal flavour violation, two common soft masses $M_{\tilde{l}_L}$ and $M_{\tilde{l}_R}$ for left-handed and right-handed sleptons, equality of the soft squark masses between the first and second generations, $M_{\tilde{q}_{1,\;2}}$, while the mass of the third generation squarks is kept as an independent free parameter $M_{\tilde{q}_3}$. We allow for only one non-zero trilinear coupling, $A_t$. The gaugino masses $M_1, M_2$ and $M_3$ are free parameters as well. In particular this allows to have $M_1\ll M_2$, implying a light neutralino much below the EW scale. The ratio of the doublet Higgs VEV' $\tan\beta$, is also a free parameter in both models. In the MSSM, the Higgs bilinear term, $\mu$, and the pseudoscalar mass $M_A$ are the remaining free parameters. Thus we consider MSSM scenarios with the 11 following free 
parameters~\footnote{We do not perform a thorough exploration of the much studied CMSSM because it is only a 
particular case of the MSSM. Furthermore in that model the LSP mass is above 50 GeV because of the LEP limit 
on charginos and the relation between the chargino and the neutralino masses. 
This leaves little possibilities for parameter determination.} 
\begin{equation}
M_1, \; M_2, \; M_3, \; \mu, \; \tan\beta , \; M_A, \nonumber\\
M_{\tilde{l}_L}, \; M_{\tilde{l}_R}, \; M_{\tilde{q}_{1,\;2}}, \; M_{\tilde{q}_3}, \; A_t. \nonumber
\end{equation}
 
The NMSSM is a simple extension of the MSSM with an additional gauge singlet superfield, S, that provides a solution to the naturalness problem. Indeed the parameter $\mu=\lambda \langle S \rangle$ is determined by the VEV of the scalar singlet and is thus naturally of the EW scale~\cite{Ellwanger:2009dp}. In the NMSSM there are additional parameters related to the extended Higgs sector. The part of the superpotential involving Higgs fields reads
\begin{equation}
W=\lambda S H_uH_d +\frac{1}{3} \kappa S^3 \nonumber
\end{equation}
and the soft Lagrangian is
\begin{eqnarray}
{\cal L}_{\rm soft} = m^2_{H_u}|H_u|^2 + m^2_{H_d}|H_d|^2+m^2_{S}|S|^2\ \nonumber\\
+(\lambda A_\lambda H_u H_d S+ \frac{1}{3} \kappa A_\kappa S^3+h.c.).\nonumber
\end{eqnarray}
After using the minimization conditions of the Higgs potential, the Higgs sector, which consists of three neutral scalar fields, $H_1,\; H_2,\; H_3$ and two pseudoscalar neutral fields, $A_1,\; A_2$ as well as a charged Higgs, $H^\pm$ is described by six free parameters, $\mu$, $\tan\beta$ as well as $\lambda$, $\kappa$, $A_\lambda$ and $A_\kappa$. The list of free parameters therefore contains the ones of the MSSM with the pseudoscalar mass, $M_A$, replaced by 
\begin{equation}
\lambda, \kappa, A_\lambda, A_\kappa ,\nonumber
\end{equation}
for a total of 14 free parameters.

These simplified models reproduce the salient features of neutralino DM. Indeed, apart from the mass of the LSP, the most important parameters are the gaugino/higgsino content of the LSP, determined by $\mu$ and $M_1$, $M_2$, $\tan \beta$, as well as the mass of the Higgses which can enhance significantly neutralino annihilations. Sfermion exchange, and in particular slepton exchange, can also play a role for light neutralinos. 

There are many similarities between the MSSM and the NMSSM, as will be seen in the following analysis. However one characteristic feature of the NMSSM is that the singlet fields, which mostly decouple from the SM fields, can be very light and yet escape LEP bounds~\cite{Ellwanger:2009dp}. Therefore it is much easier to have light neutralinos because 
they can annihilate into or through the exchange of light singlet Higgses~\cite{Belanger:2005kh}.
%fm: plus leger ?  %gb: plus leger si c'est dans l'etat final mais 2mneut si echange en canal-s

\subsection{Scanning method} \label{sec:Meth}

In order to thoroughly scan the parameter space we used a Markov Chain Monte-Carlo (MCMC) code, first presented in~\cite{Vasquez:2010ru}. The scanning procedure consists on a Metropolis-Hastings algorithm, and is based on micrOMEGAs2.4~\cite{Belanger:2007zz,Belanger:2008sj,Belanger:2010gh} for the computation of all observables. The supersymmetric spectra are calculated with SuSpect~\cite{Djouadi:2002ze} in the MSSM and with NMSSMTools~\cite{Ellwanger:2005dv} in the NMSSM. The latter also provides collider constraints on the Higgs sector, on sparticles and on flavour observables. 

Each point is generated by making a random step with a normal variation from the previous point in each dimension. Then, we compute its total prior $\mathcal{P}$, total likelihood $\mathcal{L}$ and total weight $\mathcal{Q}=\mathcal{P}\times\mathcal{L}$. It is kept with a probability $Min\left(1,\; \mathcal{Q}'/\mathcal{Q}\right)$, where $\mathcal{Q}'$ is the total weight of the point being tested and $\mathcal{Q}$ is that of the source point. If the evaluated point is not kept, then a new point is generated from the last accepted point. Thus, the parameter space is scanned via a random walk by iterating this procedure.

The priors we impose are: a set of parameters has to lie within the boundaries of the parameter space given by Table~\ref{tab:par_int}, while a physical solution of the spectrum calculator and a neutralino LSP are required. Regarding likelihoods, these are displayed in Table I of~\cite{Vasquez:2010ru}. We include limits on B physics observables, on the anomalous magnetic moment of the muon $(g-2)_{\mu}$, on the Higgs and sparticles masses obtained from LEP and the corrections to the $\rho$ parameter. In the MSSM, the limits on the Higgs mass were applied by making use of the SUSY-HIT~\cite{Djouadi:2006bz} and the HiggsBounds packages~\cite{Bechtle:2008jh,Bechtle:2011sb} as in~\cite{Vasquez:2011yq}. The HiggsBounds version used (3.1.3) include LEP and Tevatron results as well as first LHC results, more recent results from CMS presented in~\cite{Chatrchyan:2011nx,CMS_higgs_fit} were added a posteriori. 
 Notice that we take the WMAP measurement on the DM relic density as a strict upper limit on the LSP relic density --obtained via the usual freeze out mechanism--, however, we allow the neutralino to have a relic density as low as $10\%$ of the measured value. Indeed, the LSP could be only a fraction of the dark component, the rest corresponding to other dark particles or to a modified theory of gravity. For more details see~\cite{Vasquez:2010ru}.

The scans we performed in this study were aimed at giving a general determination of the different configurations with neutralino masses at the weak scale and below. However, as it was shown in~\cite{Vasquez:2010ru,Vasquez:2011yq}, it is difficult to find light ($\lesssim 30$ GeV) neutralinos with a random walk in both the MSSM and the NMSSM: the probability of falling in these regions that require fine-tuning is  rather small. Indeed, in the MSSM, the neutralino LSP has to annihilate via the exchange of either rather light Higgs bosons, scenarios that are heavily constrained by the Tevatron experiments as well as by CMS, or light sleptons, particularly of staus with masses close to the LEP lower bound of 81.9 GeV~\cite{Nakamura:2010zzi}. In the case of the NMSSM, it has been shown in~\cite{Vasquez:2010ru,Vasquez:2011js} that the neutralino can achieve light masses by annihilating via or into very light singlet-like Higgs bosons.

Hence we used two different techniques to trigger the chains that scanned the parameter spaces. On one hand we let part of the chains start randomly, {\it i.e.}, look randomly for a starting point with $\mathcal{Q}\neq 0$. On the other hand we used the previous knowledge of fine-tuned regions explored in~\cite{Vasquez:2010ru,Vasquez:2011yq} to set fixed starting points for the rest of the chains, in order to force the random walk to yield at least a few points in such regions.

A summary of the characteristics of the runs we present is given in Table~\ref{tab:runs}.

\begin{table}[hbt]
\centering
\begin{tabular}{|c|c|c|c|c|}
\hline
&&&& \\
\rm{Parameter} & \rm{Minimum} & \rm{Maximum} & \rm{Tolerance} & \rm{Model} \\
&&&& \\
\hline
$M_1$ & 1 & 1000 & 3 & both \\
$M_2$ & 100 & 2000 & 30 & both \\
$M_3$ & 500 & 6500 & 10 & both \\
$\mu$ & 0.5 & 1000 & 0.1 & both \\
$\tan\beta$ & 1 & 75 & 0.01 & both \\
$M_A$ & 1 & 2000 & 4 & MSSM \\
$\lambda$ & 0 & 0.75 & 0.1 & NMSSM \\
$\kappa$ & 0 & 0.65 & 0.08 & NMSSM \\
$A_\lambda$ & -5000 & 5000 & 100 & NMSSM \\
$A_\kappa$ & -5000 & 5000 & 100 & NMSSM \\
$A_t$ & -3000 & 3000 & 100 & both \\
$M_{\tilde{l}_R}$ & 70& 2000 & 15 & both \\
$M_{\tilde{l}_L}$ & 70 & 2000 & 15 & both \\
$M_{\tilde{q}_{1,\,2}}$ & 300 & 2000 & 14 & both \\
$M_{\tilde{q}_3}$ & 300 & 2000 & 14 & both \\
\hline
\end{tabular}
\caption{Intervals of free parameters used for the MSSM and NMSSM scans (GeV units).}
\label{tab:par_int}
\end{table}

\begin{table}[hbt]
\centering
\begin{tabular}{|c|c|c|c|c|c|}
\hline
& \rm{Points} & $\mathcal{Q}_{max}$ & $1\,\sigma$ & $2\,\sigma$ & $3\,\sigma$ \\
\hline
\hline
MSSM & 1208949 & 0.755 & 0.25 & 0.68 & 0.97 \\
\hline
NMSSM & 2092875 & 0.812 & 0.30 & 0.72 & 0.98 \\
\hline
\end{tabular}
\caption{Basic characteristics of the scans of the MSSM and the NMSSM. The $1\,\sigma$, $2\,\sigma$ and $3\,\sigma$, columns represent the fraction of points satisfying $0.32\times\mathcal{Q}_{max}\leq \mathcal{Q}$, $0.05\times \mathcal{Q}_{max} \leq \mathcal{Q}$ and $0.003\times \mathcal{Q}_{max} \leq \mathcal{Q}$ respectively.}
\label{tab:runs}
\end{table}

\section{Results}
\label{sec:results}
Before analyzing the predictions for SD interactions, we impose further constraints on the parameter space of the models found by the MCMC. We focus on astroparticle constraints, including XENON100 limits on SI interactions~\cite{Aprile:2011hi} as well as limits from Fermi-LAT observations of the photon flux from dSphs~\cite{Abdo:2010ex}\footnote{Here we do not show the impact of recent LHC limits on the MSSM Higgs sector, in particular, 
the LHCb and CMS combined results on the $B_s\rightarrow\mu\mu$ branching ratio and the CMS negative search 
for  $H\rightarrow \tau\tau$  constraining the $\tan\beta$ vs. $M_A$ plane. For an account of such an analysis 
over the same MSSM data set presented here, see~\cite{AlbornozVasquez:2011qc}.}.

\subsection{Dark Matter searches constraints}\label{sec:astropart}
Dark Matter observables are computed for each point kept by the MCMC. This includes the direct detection cross sections: SI and SD neutralino-nucleon elastic scattering processes, for both protons and neutrons, and the $\gamma$-ray flux produced by neutralino annihilations at low velocities.

Notice that the actual observables are:  
$\sigma_{scat} \rho_\odot$ for direct detection, {\it i.e.} the elastic scattering cross section times the neutralino density at Earth, 
and $\sigma_{ann} \rho^2_{loc}$ for $\gamma$-ray indirect detection, {\it i.e.} the  neutralino annihilation cross section times the local neutralino density squared
(in the astrophysical object). Since we allow for the neutralino to represent only a fraction of the DM component, we scale the neutralino density by the same fraction in all astrophysical systems. Thus we define
\begin{align}
\xi = Min\left(\frac{\Omega_{\chi_1^0}h^2}{\Omega_{WMAP}h^2},\;\; 1\right)\nonumber
\end{align}
where $\Omega_{WMAP}h^2=0.1097$ is the $1\sigma$ lower limit of the DM density as obtained by the WMAP 5-year analysis~\cite{Komatsu:2008hk}.

\subsubsection{Spin Independent elastic scattering}
Spin independent interactions result from Higgs and/or squark exchanges. Because of the mass scales involved, it is generally 
the Higgs contribution that dominates, providing the LSP is some mixture of higgsino/gaugino. In the MSSM, both the light and heavy Higgses can contribute, in particular the heavy Higgs contribution is enhanced at large values of $\tan\beta$. In the NMSSM, there is in addition a contribution 
from the singlet Higgs. When the lightest scalar is below 10 GeV one can have an enhancement of the cross section even if the LSP is weakly coupled to the scalar Higgs~\cite{Vasquez:2011js}. The SI interaction rates depend on the quark content of the nucleons. Indeed, the choice of quark coefficients in nucleons could vary the estimation by an order of magnitude. In micrOMEGAs the quark content of nucleons is parametrized by the $\sigma_{\pi N}$ and $\sigma_0$ terms~\cite{Belanger:2008sj}. Recent lattice QCD results point towards small strange quark contributions, hence to $\sigma_{\pi N}\simeq \sigma_0$~\cite{Giedt:2009mr}. We take $\sigma_{\pi N}=45$~MeV and $\sigma_0=40$~MeV.

\begin{figure}[bt]
\centering
\includegraphics[width=0.5\textwidth ,natwidth=6cm,natheight=6cm]{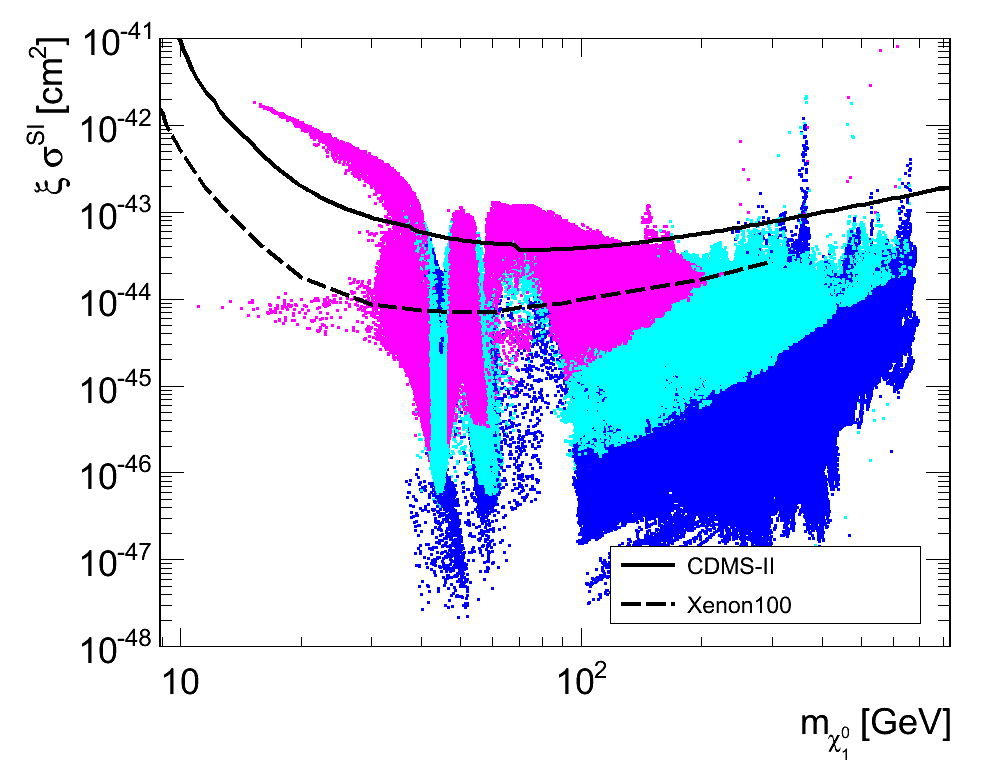}
\includegraphics[width=0.5\textwidth ,natwidth=6cm,natheight=6cm]{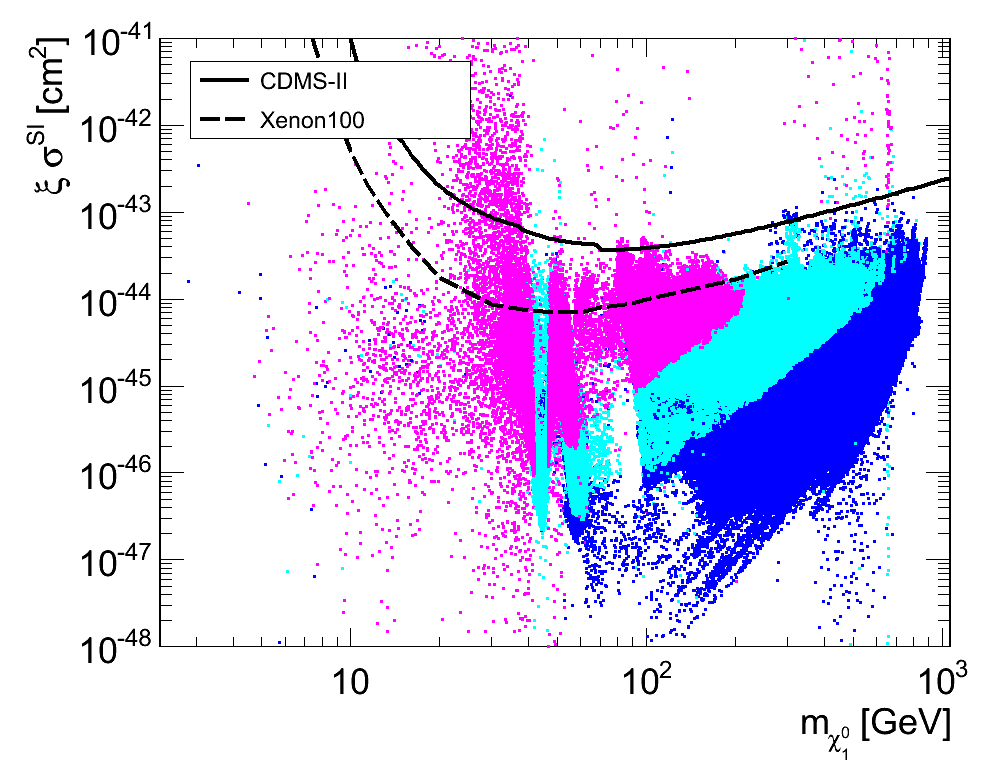}
\caption{Spin independent cross section versus the neutralino mass. Top: MSSM. Bottom: NMSSM. In pink the points in the discovery region of directional detectors and in cyan in the exclusion region. The CDMS-II~\cite{cdms-spin} and XENON100~\cite{Aprile:2011hi} limits are also displayed.}
\label{fig:SI_vs_Mchi_MIMAC}
\end{figure}

The predictions for SI interactions in the MSSM and the NMSSM along with XENON100~\cite{Aprile:2011hi} and 
CDMS-II~\cite{cdms-spin} limits are displayed in Fig.~\ref{fig:SI_vs_Mchi_MIMAC}. Color tagging in relation with 
detectability in directional detection will be discussed in Sec.~\ref{sec:complement}. The XENON100 limit attain some
 of the configurations found in both models, thus constraining the parameter spaces. Nevertheless, most scenarios lie below  
 the XENON100 sensitivity, and many are more than an order of magnitude away from it. This highlights the need for complementarity in DM
 search strategies.

While in general the point distribution is rather similar in both models, there are a few differences between the two panels. 
Those differences can best be understood by investigating the effet of the relic density constraint  which depends on the model.
In particular, neutralinos below 30 GeV in the NMSSM usually imply a light Higgs scalar or pseudoscalar to 
ensure the annihilation rate is large enough to obtain  relic densities below the WMAP value. If the light Higgs  is a scalar, then the SI cross section gets enhanced, as  shown in~\cite{Vasquez:2011js,arXiv:1009.3963}, whereas a light pseudoscalar does not contribute to the SI cross section. This explains the larger range of predicted cross sections  observed in the NMSSM with respect to the MSSM. In the latter, as discussed in~\cite{Vasquez:2011yq}, when the relic density is achieved through as-light-as-possible scalar Higgs exchanges, the lighter the neutralino, the wider the mass difference between neutralinos and Higgses, thus the larger the coupling between these must be. Hence the SI cross section tends to increase towards lighter neutralino masses, which corresponds to the top-left arm of the cloud in Figs.~\ref{fig:SI_vs_Mchi_MIMAC}'s top panel. However, when the neutralino annihilations are driven by slepton exchanges, the SI cross section does not evolve with the neutralino mass, and is not necessarily enhanced by the need of large neutralino-Higgs couplings. That situation corresponds to the other light neutralino arm, a more diffuse cloud at smaller interaction rates, in the same panel.

Above 30 GeV, neutralinos can achieve the correct relic density by resonant Z exchange. That region (for $m_{\chi_1^0} \sim 45$ GeV) is well represented in both models. In both cases many points show small SI rates. Since the Z resonance enhances the annihilation rate, the Z$\chi_1^0\chi_1^0$ coupling should be suppressed, in order to have enough neutralinos left after the thermal freeze-out.   Hence the Higgs couplings, which depend on the higgsino component of the LSP as the Z coupling, are also suppressed, which in turn diminishes the SI cross section. Above the Z resonance, and around 60 GeV, the SM-like Higgs exchange on resonance dominates the relic density. For the same reason, when the masses are  very fine-tuned, $m_{\chi_1^0}\approx m_h/2$,  the required annihilation rate   forces the Higgs-neutralino coupling (responsible for the interaction) to be fairly small, therefore allowing very weak SI interactions. Note that the 45-85 GeV neutralino mass range with large cross-sections is more populated in the MSSM than the NMSSM. 
%This could be understood in terms of the constrained mass of the SM-Higgs in the MSSM, which is most likely to lie between 110 and 130 GeV, while in the NMSSM more freedom is allowed due to the extra singlet in the Higgs sector.
This could be understood in terms of the couplings of the  SM-like Higgs to the LSP which 
can be suppressed because of the singlet component in the NMSSM.

Approaching 100 GeV, neutralinos could have larger higgsino components, since we can have $\mu \gtrsim 100 GeV \sim m_{\chi_1^0}$. Furthermore there are many possibilities for  neutralinos to achieve the relic density: by Z or Higgs exchanges (large higgsino components) or  via sfermion and gaugino exchanges. 
 Also, the heavier the neutralino, the narrower the neutralino-squark mass difference could be. This enchances the squark contribution and allows for some unusually large cross sections. The predictions for the SI cross section span five orders of magnitute. In general, the heavier the neutralino, the smaller the cross section and the XENON100 limits eventually do not constrain many configurations with $m_{\chi_1^0}<200$~GeV.

\subsubsection{Gamma rays}
Neutralinos in galactic objects, such as the Milky Way and its dwarf spheroidal companions (hereafter referred to as dSph), have a probability of encountering and annihilating into SM fermions. After the subsequent decays and hadronization, these events produce $\gamma$-rays. More marginally, $\gamma$-rays can be produced from  internal lines in the annihilation process 
%from final state radiation, 
or directly by pairs, the latter which occur through a loop-induced process  is not included in our flux computation. Among the indirect signatures, $\gamma$ rays can be easily computed, as compared to charged cosmic rays,   they do not suffer from energy losses and their propagation is not subject to uncertainties on the interstellar medium. Also, the Fermi-LAT, HESS and MAGIC experiments are successfully exploring the fluxes at Earth, thus providing means to constrain DM models.

The expected flux of $\gamma$ photons of energy $E$ for a given angular spread $\psi$, $\phi _{\gamma} \left(E, \psi\right)$, depends upon the square of the neutralino density, on the annihilation cross section $\left< \sigma v \right>$ times the squared local abundance of neutralinos $\rho^2_{\chi_1^0}$ and on the particle spectrum produced $\frac{dN\left(E\right)}{dE}$. The computation of the flux implies the integral over the line of sight (l.o.s.) $l$ within the extension of the desired area, which is usually matched to that of an observatory. Hence
\begin{align}
\phi _{\gamma} \left(E, \psi\right) =& \nonumber \int_{l.o.s.} dl\left(\psi\right)\rho^2_{\chi_1^0}\left(l\left(\psi\right)\right) \, \times \, \frac{1}{2}\frac{\left< \sigma v \right>}{4\pi m_{\chi_1^0}^2} \frac{dN\left(E\right)}{dE} \nonumber
\\
=& J\left(\psi\right) \times \phi^{PP}\left(E\right). \nonumber
\end{align}
We have explicitly split the so called astrophysics term $J\left(\psi\right)$ --namely the l.o.s. integral containing all the terms which carry spatial dependence-- and particle physics term $\phi^{PP}\left(E\right)$ --which is the energy dependent term. The former can only be determined with the knowledge of the DM distribution in the observed object. 
We focus here on dSphs observed by Fermi-LAT~\cite{Abdo:2010ex}, for which they provide an estimation of $J\left(\psi\right)$ (see their Table 4). We compute $\int \phi^{PP}\left(E\right)$ over the $\left[100\,\rm MeV,\; m_{\chi_1^0}\right]$ interval. Here, we have scaled down the flux by a $\xi^2$ factor in order to take into account only the neutralino contribution to the DM density distribution.

\begin{figure}[bt]
\centering
\includegraphics[width=0.5\textwidth ,natwidth=6cm,natheight=6cm]{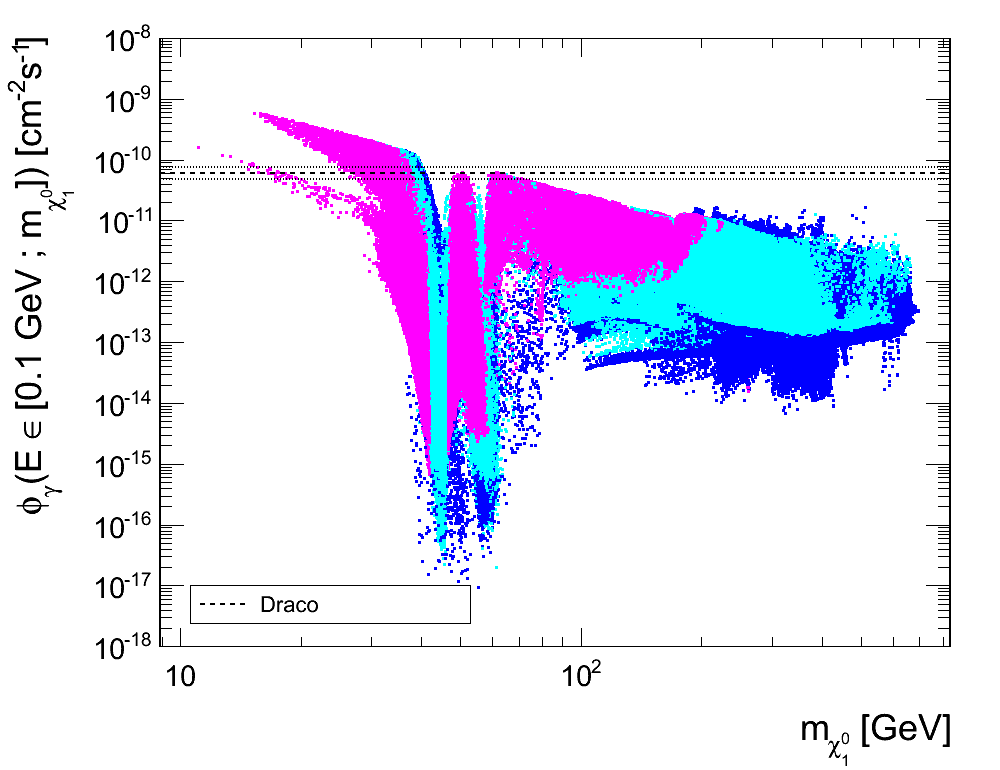}
\includegraphics[width=0.5\textwidth ,natwidth=6cm,natheight=6cm]{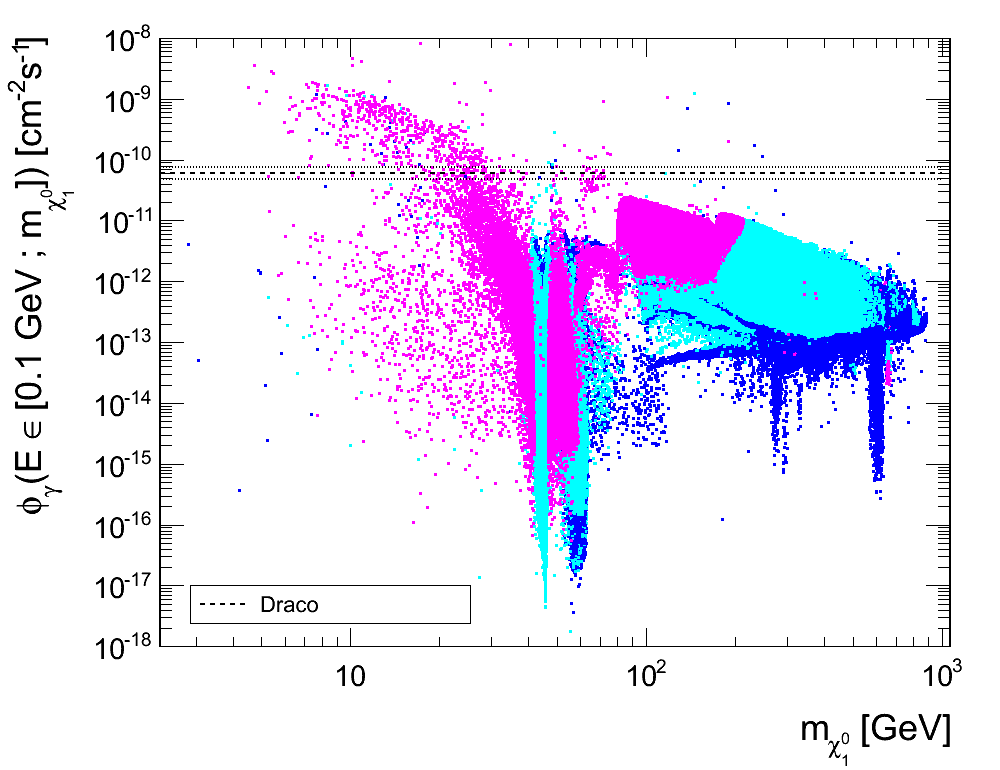}
\caption{Flux of $\gamma$-rays expected from neutralino annihilations from the Draco dwarf spheroidal galaxy versus the neutralino mass. Top: MSSM. Bottom: NMSSM. In pink the points in the discovery region of directional detectors and in cyan in the exclusion region. The Fermi-LAT limits~\cite{Abdo:2010ex} are also displayed.}
\label{fig:Dwarf_vs_Mchi_Draco_MIMAC}
\end{figure}

The annihilation cross section at galactic velocities is related to the interaction rate at freeze out. However, the velocity in galaxies is much lower than at freeze out, hence large variations can take place when the annihilation occurs via resonances  in the early universe~\cite{Griest:1990kh}. 
In general annihilations that proceed through a Higgs and/or Z resonance in the early universe are diminished at $v\sim 10^{-3}c$, furthermore sfermion exchange into light fermions get suppressed. However it is also possible to have an enhancement of the neutralino annihilation cross section in galaxies when it is dominated by the exchange of a Higgs with a narrow width, see~\cite{Vasquez:2011js} for a discussion on this subject in the particular case of the light neutralinos in the NMSSM.

Fig.~\ref{fig:Dwarf_vs_Mchi_Draco_MIMAC} shows the $\gamma$-ray integrated flux expected from the Draco dSph as a function of the neutralino mass, along with the Fermi-LAT limits~\cite{Abdo:2010ex}. Color tagging in relation with detectability in directional detection will be discussed in Sec.~\ref{sec:complement}. Fewer points are challenged by Fermi-LAT than by XENON100. Only the lighter configurations can be constrained by their $\gamma$-ray yield, while heavier neutralinos seem to be out of the reach of the Fermi-LAT detector. 
As for SI interactions, results for the MSSM and the NMSSM differ mainly for the lighter neutralino configurations. Here, 
the scalar Higgs exchanges suffer from a low-velocity suppression, 
while the pseudoscalar exchanges can be resonantly enhanced, this is possible only  in the NMSSM. 
Thus, again, the NMSSM predicts a broader range of possible $\gamma$-ray yields when neutralinos are light. An order of magnitude enhancement of the Fermi-LAT sensitivity would probe all the configurations in the MSSM up to 30~GeV neutralino masses, and many scenarios up to 200~GeV in both models.

\subsection{Predictions for Spin Dependent interactions}\label{sec:SD}

\begin{figure}[bt]
\centering
\includegraphics[width=0.5\textwidth ,natwidth=6cm,natheight=6cm]{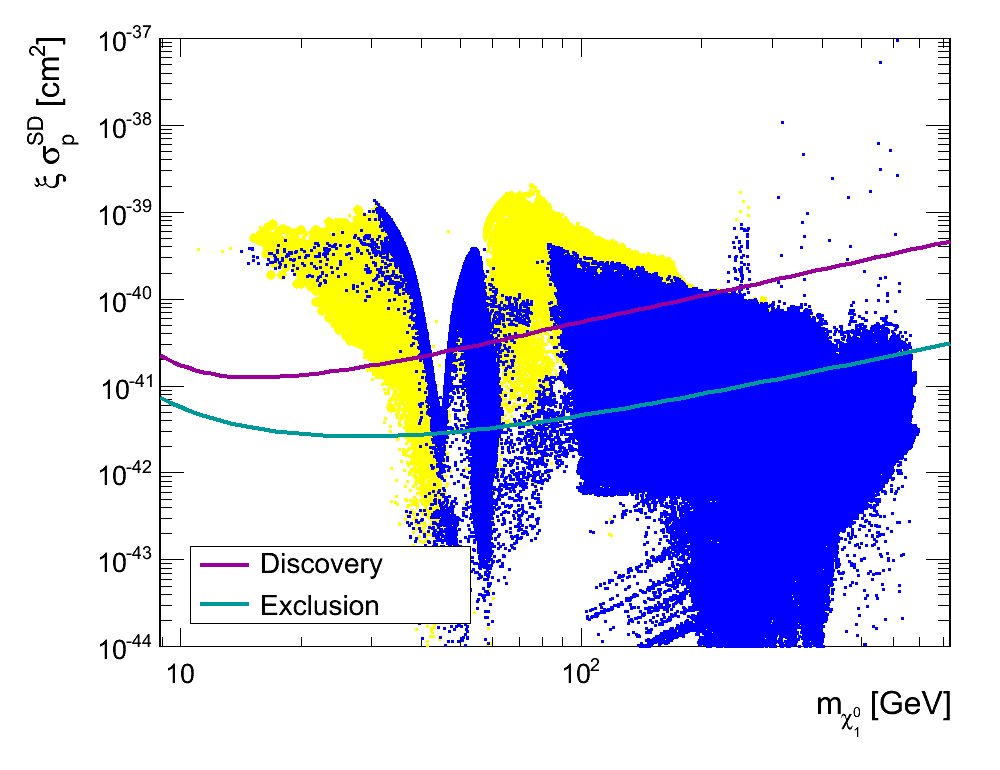}
\includegraphics[width=0.5\textwidth ,natwidth=6cm,natheight=6cm]{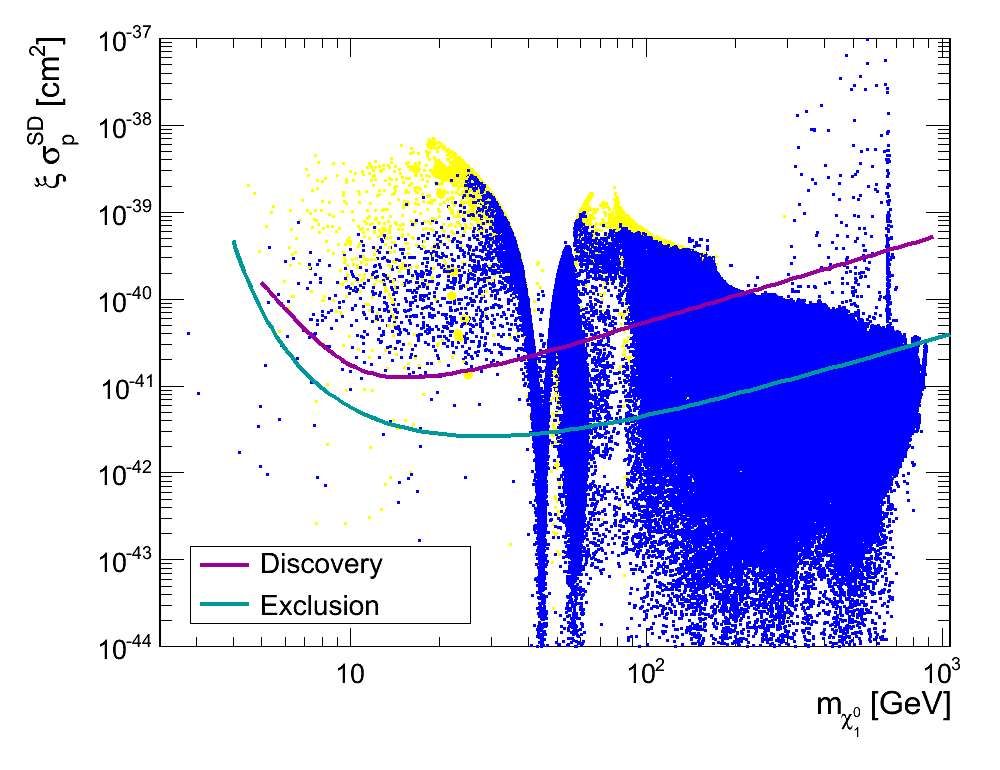}
\caption{Proton-neutralino spin dependent elastic scattering cross section versus the neutralino mass with the exclusion and discovery projections for a nominal directional detector. Top: MSSM. Bottom: NMSSM. In blue safe points and in yellow points excluded by either XENON100 or Fermi-LAT.}
\label{fig:SDp_vs_Mchi_MIMAC}
\end{figure}
%fm deja dit 3 fois...
%Taking advantage on the rotation of the Solar system around the galactic center through the DM halo, 
%directional detection should allow to show a direction dependence of WIMP events.
%As emphasized in section~\ref{sec:directional}, this can be used, within the framework of dedicated data analysis, 
%either to exclude a given WIMP model, defined by the mass and cross section, or to claim a high significance discovery, the main incoming
%direction being the proof of discovery. It is worth noticing however that low pressure directional detector strategy implies to focus on 
%spin-dependent  interaction to be competitive with planned ton-scale direct detectors. 

The predictions for the $\xi \sigma_p^{SD}$ observable, which is the target of the future directional detectors, 
can be seen in Fig.~\ref{fig:SDp_vs_Mchi_MIMAC}. We also show the projected curves for the discovery limit 
(in pink), defined as  a significance
greater than the 3$\sigma$ level at 90\% Confidence Level (C.L.) in 30 kg.year $CF_4$ directional detector, 
and the exclusion limit (in cyan), in the background-free case for the same detector with the same exposure, see sec.~\ref{sec:directional}.  These curves define three regions: 
the discovery region above the discovery curve, the exclusion region between the two curves, 
and the out-of-reach region below the exclusion limit. We have tagged  points failing to overcome the XENON100 and/or the Fermi-LAT constraints in yellow. MSSM and NMSSM configurations range over several orders of magnitude, the maximum reaches $10^{-39}$ cm$^2$ and a large fraction of the points lie above the potential exclusion limit of future directional detectors. One can readily see that some configurations that are not yet constrained by SI detectors nor by indirect signals (blue points)
 lie in the exclusion or even the discovery region.

\begin{figure}[bt]
\centering
\includegraphics[width=0.5\textwidth ,natwidth=6cm,natheight=6cm]{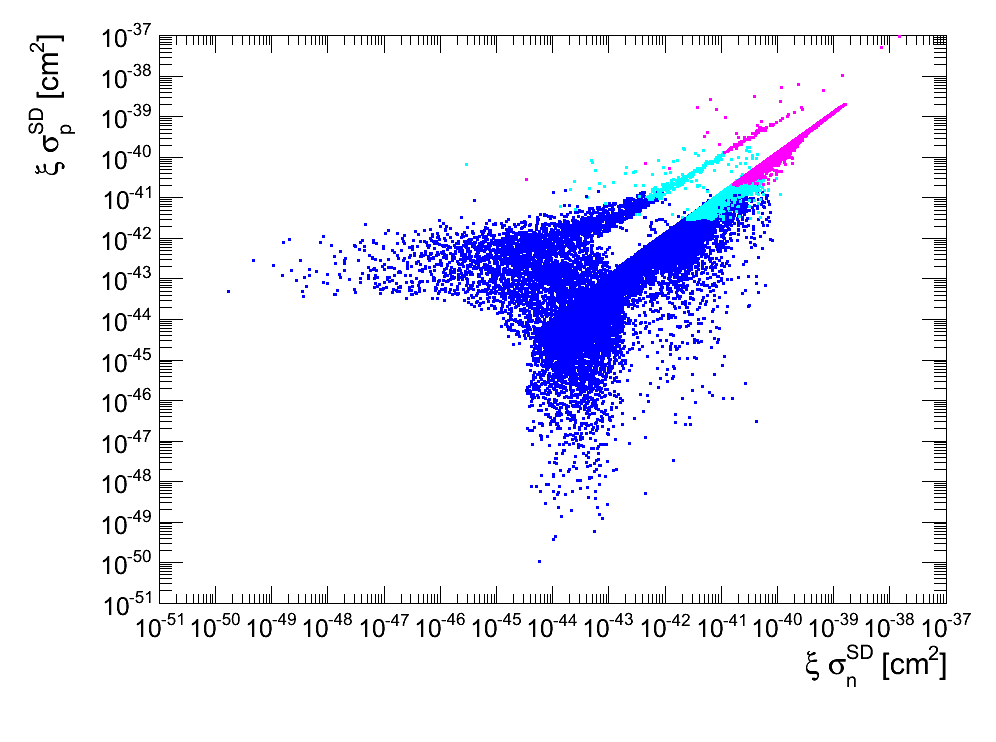}
\includegraphics[width=0.5\textwidth ,natwidth=6cm,natheight=6cm]{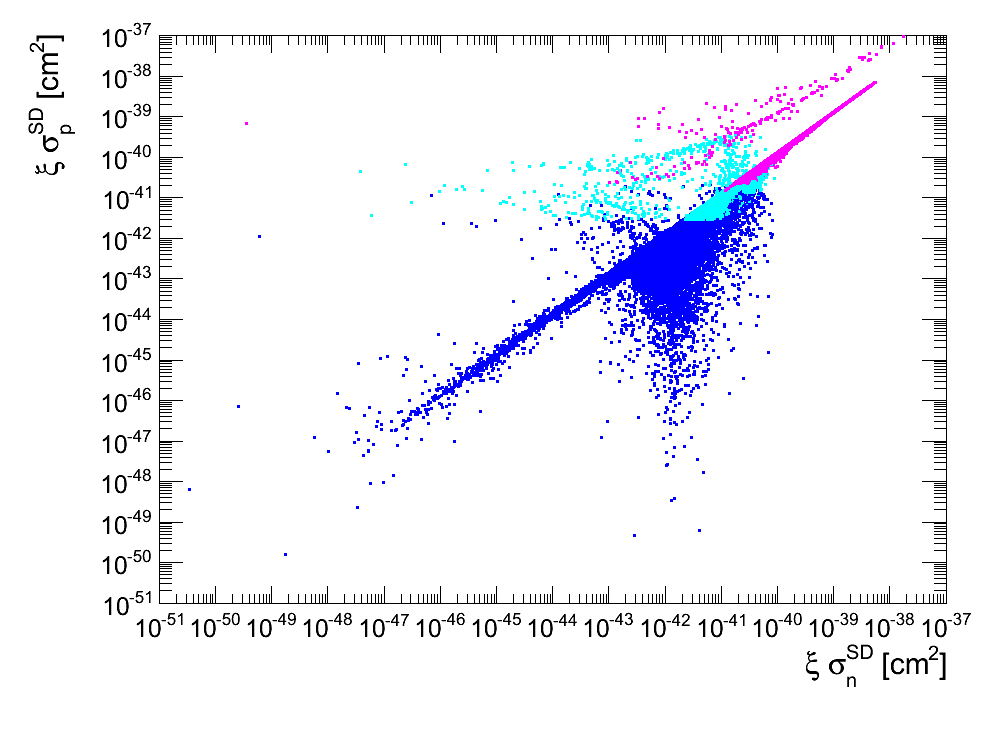}
\caption{Spin dependent elastic scattering cross sections correlations: proton-neutralino versus 
neutron-neutralino interactions. Contrarily to fig.~\ref{fig:sigsig-all}, the result is presented for all neutralino masses. 
Top: MSSM. Bottom: NMSSM. In pink the points in the discovery region of directional detectors and in cyan in the exclusion region.}
\label{fig:SDp_vs_SDn_MIMAC}
\end{figure}

\begin{figure}[bt]
\centering
\includegraphics[width=0.5\textwidth ,natwidth=6cm,natheight=6cm]{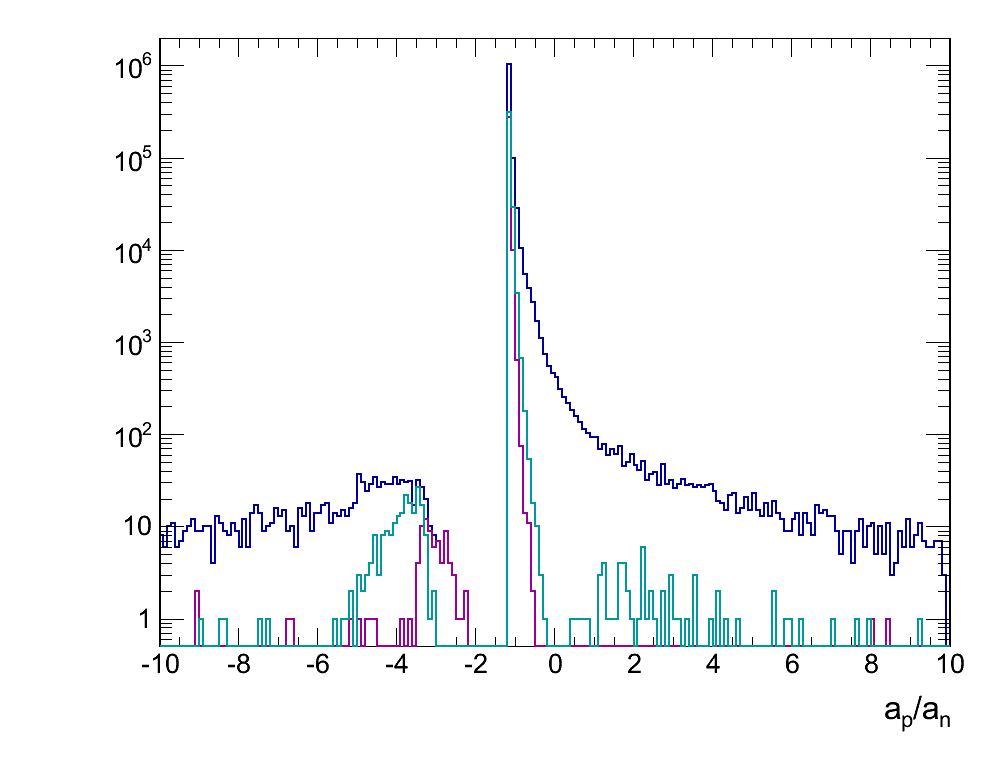}
\includegraphics[width=0.5\textwidth ,natwidth=6cm,natheight=6cm]{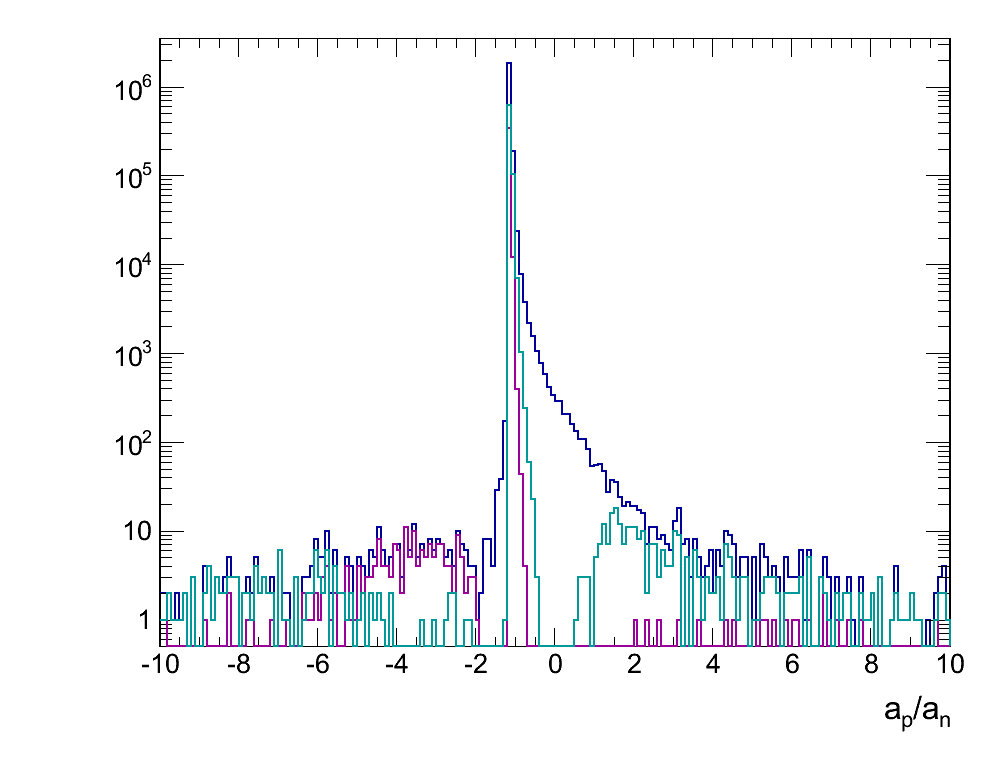}
\caption{Frequency distribution of $ap/an$. 
Top: MSSM. Bottom: NMSSM. In pink the points in the discovery region of directional detectors and in cyan in the exclusion region.}
\label{fig:ASDp_sur_ASDn_MIMAC}
\end{figure}

In particular most neutralinos with a mass $m_{\chi_1^0}\leq 40$ GeV lead to large cross sections. This is understandable since in this case the same diagram (Z exchange) is responsible for both annihilation and scattering on nucleon. Enhancing the annihilation cross section in order to get an acceptable relic density implies enhancing the SD cross sections. However, when the neutralino mass approaches $m_Z/2$, the Z-neutralino coupling is reduced so the relic density is not too small, hence yielding very small SD interaction rates.

For larger neutralino masses, the possible cross section values range similarly to the SI case. 
However, here there is a clear upper limit for the SD, Z-exchange preferred configurations. 
For $m_{\chi_1^0}\geq100$ GeV the SD cross section scales as $1/m^2_{\chi_1^0}$. 
Also, the largest coupling between the Z and the neutralino is  $Max(N_{13}^2-N_{14}^2)=0.5$, 
which can be met for higgsino dominated neutralinos. Those configurations having the maximum coupling describe the upper limit of the cloud above 70~GeV, drawing the $1/m^2_{\chi_1^0}$ curve. 
Points falling above that limit imply dominant squark exchanges, and points below have either smaller 
couplings or destructive interference between the Z and squark exchange diagrams.\\
%fm
It is worth emphasizing that  in the MSSM, and to a lesser extent in the NMSSM,  a large fraction of the supersymmetric configurations 
below $m_{\chi_1^0}\leq 200$ GeV lies in the discovery region, meaning that such models could be discovered, with a significance 
greater than $3 \sigma$ (90\% CL), with a 30 kg.year $CF_4$ directional detector. Exclusions may be reached up to $\sim 800 \ GeV$.\\

%fm
As outlined in sec.~\ref{sec:xs}, a model independent constraint on DM requires to present the results 
in the ($\sigma_p^{SD}, \sigma_n^{SD}, m_\chi$) parameter 
space\footnote{Note however that it requires to set the SI coupling to zero, which is not model independent {\it stricto sensu}.}. In
order to have a complete view of the theoretical predictions, we present on Figs.~\ref{fig:SDp_vs_SDn_MIMAC} 
the $\sigma^{SD}$ on proton versus neutron, for all neutralino masses. The discovery (resp. exclusion) limit depends on the mass, 
ranging from $\sim 2 \times 10^{-41} \ {\rm cm}^2$ ($\sim 7 \times 10^{-42} \ {\rm cm}^2$) for a $\sim 10$ GeV WIMP to  
$\sim 2 \times 10^{-40}  \ {\rm cm}^2$ ($\sim 2 \times 10^{-41} \ {\rm cm}^2$)) for a $\sim 500$ GeV WIMP. 
Color tagging refers to the detectability in directional detection: pink discovery region, cyan exclusion region and blue out-of-reach.
As expected, in models dominated by Z exchange, we found $\sigma_p/\sigma_n=1.3$ (see Sec.~\ref{sec:xs1}). 
Most of the points in the discovery region (pink points) satisfy this condition. Exceptions are points where $N_{13}^2-N_{14}^2$ is small and the squark are light (more specifically below 300 GeV). The latter could easily be excluded by the LHC, unless they are degenerated with neutralinos, see Sec.~\ref{sec:LHC} for a more complete discussion on that respect.

When the squark exchange comes into play, interference between the squark and Z contributions can reduce either the neutron or proton cross sections, and the correlation is lost. Indeed, when the Z exchange is dominant but the squark exchange becomes non negligible, the proton cross section is lessened rather than the neutron cross section (points deviating from the correlation line towards the bottom). When the Z and squark exchanges are similar, then the destructive interference can completely erase one 
or the other of cross section, but not both, depending on the nature of the lightest squark. These cases correspond to the broad vertical (proton cross section suppression) and horizontal (neutron cross section suppression) distributions towards smaller cross sections in the top panel of Fig.~\ref{fig:SDp_vs_SDn_MIMAC}. Finally, when the squark exchange dominates, the proton cross section being favored, the correlation reaches $\sigma_p/\sigma_n=11.4$ (see Sec.~\ref{sec:xs1}).
In the NMSSM the correlation  $\sigma_p/\sigma_n=1.3$ extends to lower values of the cross-section. In such a case the neutralino is dominantly higgsino with a significant singlino component, furthermore  $N_{13}\approx N_{14}$ so that the Z contribution is suppressed, hence the low cross section. The annihilation of neutralinos in the early Universe is driven by the singlino component and the Higgs sector. Such scenarios are more difficult to obtain in the MSSM, furthermore the lower bound on the relic density constrains the case of a 
dominantly higgsino LSP.\\

Fig.~\ref{fig:ASDp_sur_ASDn_MIMAC} presents the frequency distribution of $a_p/a_n$ for the points
of Fig.~\ref{fig:SDp_vs_SDn_MIMAC} with the same
color code. As outlined in sec.~\ref{sec:SDCross}, the relative sign of $a_p$ and $a_n$ is a 
key issue  for SD direct searches. Indeed the interference can  be either constructive or destructive, depending of the 
sign of the spin contents of the target nucleus (see eq.~\ref{eq:xsnoyau}). For both the MSSM and the NMSSM we observe that most models give $a_p/a_n \simeq -1.13$, corresponding to a dominant Z exchange. A small number of models give $a_p/a_n \simeq -3.5$,
corresponding to a squark ($\tilde q_R$) contribution in the pure bino limit. However, amplitudes can also be of the same sign. As we shall see in section 
\ref{sec:lhc}, this is not the case when applying recent ATLAS   limits on squarks and gluinos.  

\subsection{The complementarity between directional detectors and other techniques}\label{sec:complement}

We have already shown in Sec.~\ref{sec:SD} that directional detectors will scan an important fraction of the supersymmetric parameter space. We have also shown the power of XENON100 (in direct SI detection) and Fermi-LAT (in  indirect detection experiments) to put constraints on neutralino DM configurations in Sec.~\ref{sec:astropart}. For the NMSSM case, it has been discussed in~\cite{Vasquez:2011js} that SI direct detection  and indirect detection of $\gamma$-rays constrain different light neutralino scenarios. 
In particular the light scalar Higgs case is reachable by the former, while light pseudoscalar configurations are probed by the latter.

\begin{figure}[bt]
\centering
\includegraphics[width=0.5\textwidth ,natwidth=6cm,natheight=6cm]{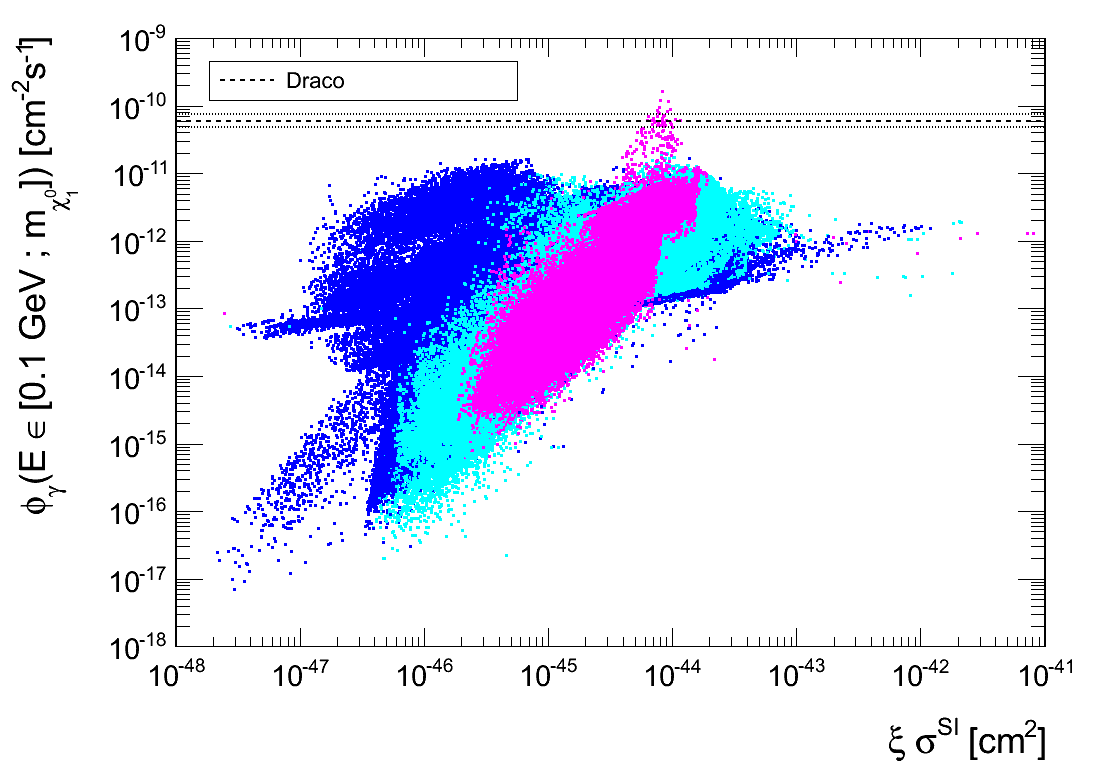}
\includegraphics[width=0.5\textwidth ,natwidth=6cm,natheight=6cm]{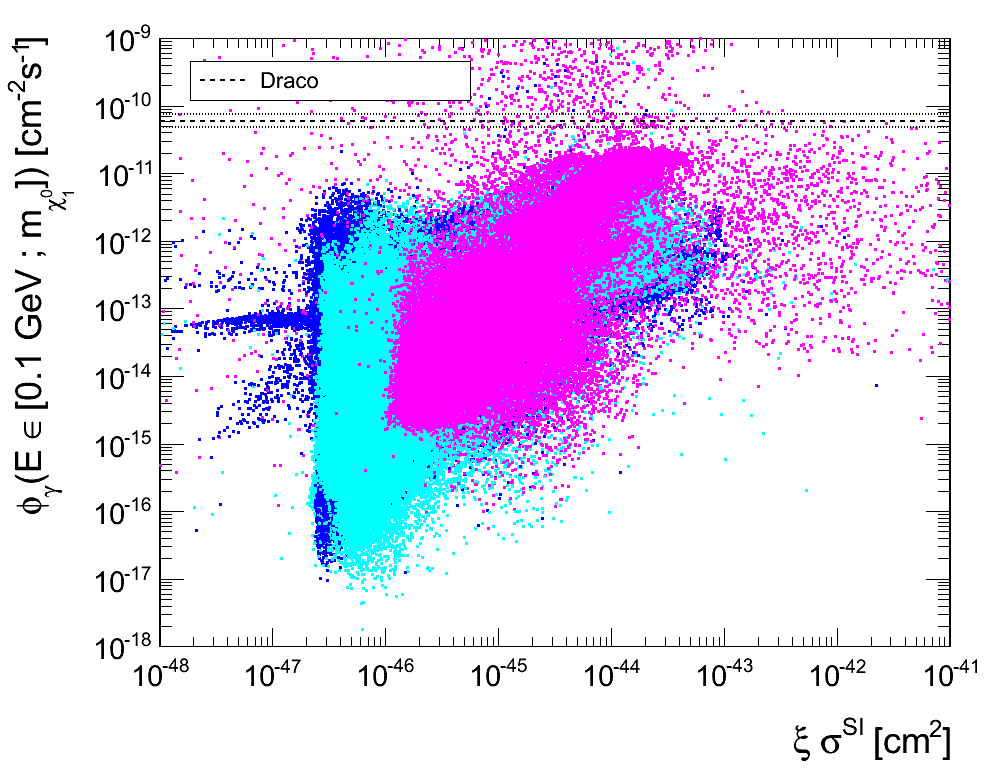}
\caption{Flux of $\gamma$-rays expected from neutralino annihilations from the Draco dwarf spheroidal galaxy versus spin independent cross section. 
Top: MSSM. Bottom: NMSSM. In pink the points in the discovery region of directional detectors and in cyan in the exclusion region. 
The Fermi-LAT limits~\cite{Abdo:2010ex} are also displayed. Points excluded by XENON100 are not shown.}
\label{fig:Dwarf_Draco_vs_SI_MIMAC}
\end{figure}

Let us discuss the impact of SD direct detection searches and their position in this sense. Figs.~\ref{fig:SI_vs_Mchi_MIMAC} and~\ref{fig:Dwarf_vs_Mchi_Draco_MIMAC} show the yields in SI cross sections and in $\gamma$-rays from the Draco dSph respectively. 
In these planes we have tagged in pink the scenarios that fall in the discovery region of future directional detectors, in cyan scenarios between the exclusion and the discovery curves, and in blue those points which are not expected to produce
 enough signal. In both cases one can see that there are many pink points below the exclusion limits of XENON100 (in Figs.~\ref{fig:SI_vs_Mchi_MIMAC}) and below those of Fermi-LAT (in Figs.~\ref{fig:Dwarf_vs_Mchi_Draco_MIMAC}). 
Most of these pink points correspond to $m_{\chi_1^0}\leq200$ GeV. We can observe that the development of a 
SD directional detector will help closing the parameter space that seems to escape other techniques. 
Unfortunately, we also observe many blue points in both figures, which lie orders of magnitude below the reach of all DM search strategies.

We summarize the complementarity of detection techniques in Figs.~\ref{fig:Dwarf_Draco_vs_SI_MIMAC}, in which we show the correlation between SI and indirect detection techniques, using the color tagging for the directional detectors potential reach. Those points failing to satisfy the XENON100 limit (which cannot be represented in this plane) are not drawn. Notice that the pink cloud, set to be upfront, covers a large region of this correlation plane, spanning over more than 2 (7) orders of magnitude in the $\xi\sigma^{SI}$ axis, and over more than 5 (7) orders of magnitude in the $\phi_\gamma$ axis for the MSSM (NMSSM). This, of course, includes many configurations that are not only far away from the current reach of SI and indirect detectors, but that would also escape their forthcoming upgrades. Supersymmetric models with neutralino DM candidates predict possible configurations in the range of detection for all the experimental techniques exposed here. However, it is obvious that none is able by itself to probe every configuration we found. Indeed, the reach of any of these techniques alone is frustratingly insufficient. In this sense, the development of SD detection techniques is useful
for a thorough scan of neutralino DM supersymmetric configurations. In particular, the superposition of the blue, 
cyan and pink clouds in Figs.~\ref{fig:Dwarf_Draco_vs_SI_MIMAC} leads us to conclude 
that the capability of SD directional detection, either to discover or to exclude a DM model, is not correlated with 
the sensitivity of the SI and indirect searches. It does in particular highlight the complementarity of the various DM search strategies.  
%increases in an uncorrelated manner with respect to the SI and indirect efforts.
%pas trop vrai pour SI, un ordre de grandeur de mieux sur toutes les masses reduit beaucoup la region pink/whatever its name is.

\subsection{Constraining MSSM and NMSSM parameter spaces with directional detectors}\label{sec:param_space}

\begin{figure}[bt]
\centering
\includegraphics[width=0.5\textwidth ,natwidth=6cm,natheight=6cm]{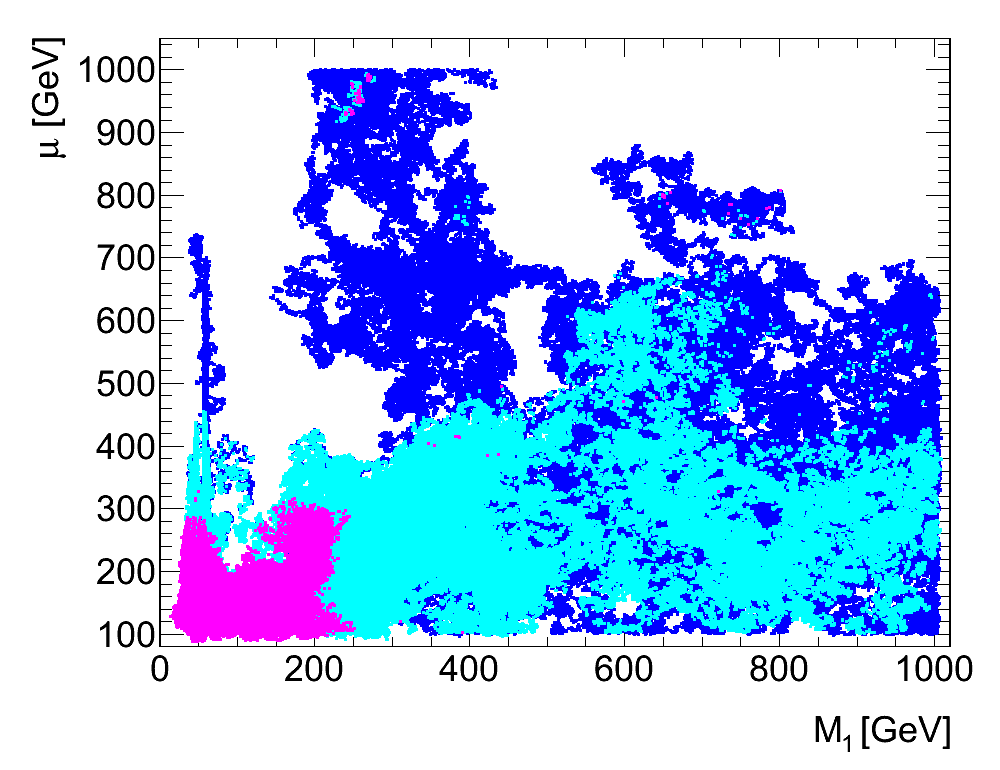}
\includegraphics[width=0.5\textwidth ,natwidth=6cm,natheight=6cm]{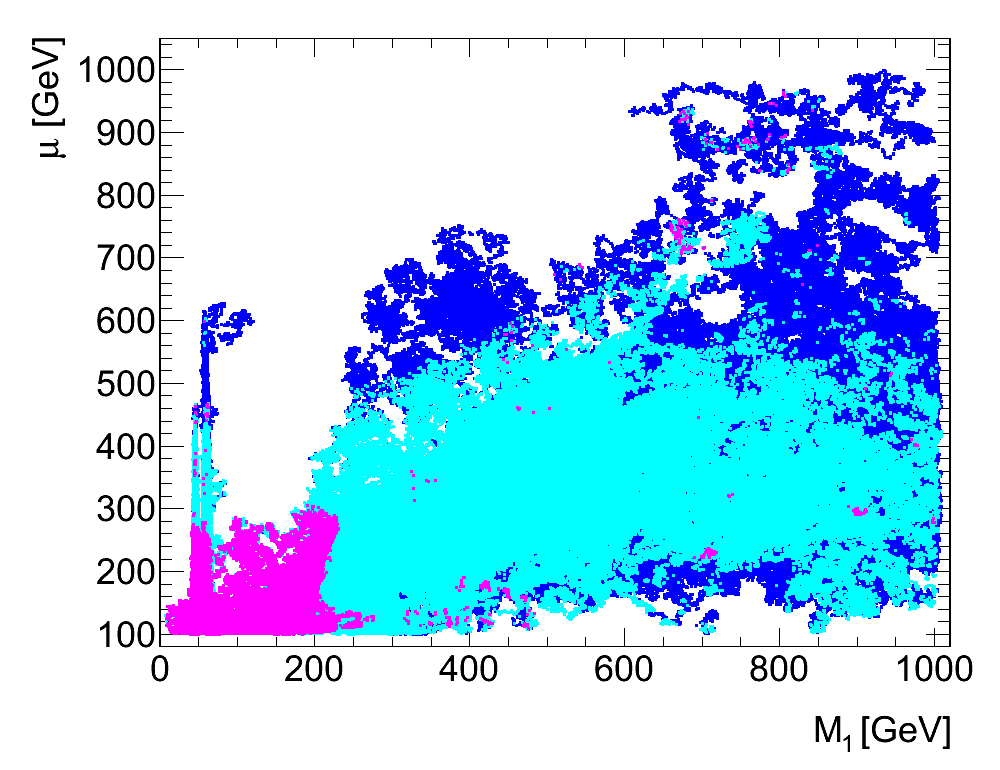}
\caption{Correlation plot between the $\mu$ mass term and the bino mass $M_1$. Top: MSSM. Bottom: NMSSM. In pink the points in the discovery region of directional detectors and in cyan in the exclusion region.} 
\label{fig:mu_vs_m1_MIMAC}
\end{figure}

\begin{figure}[bt]
\centering
\includegraphics[width=0.5\textwidth ,natwidth=6cm,natheight=6cm]{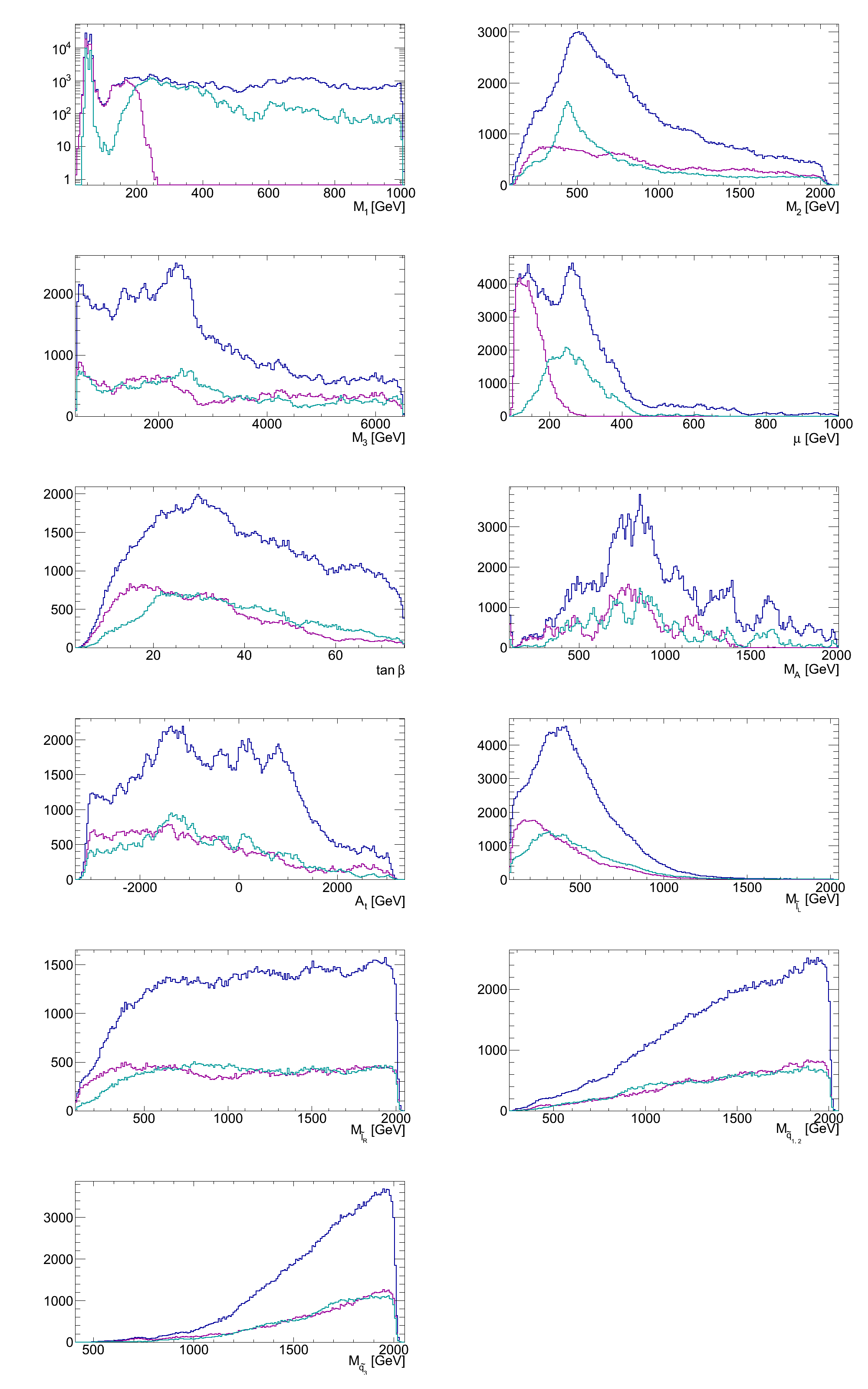}
\caption{Free parameter frequency distribution normalized to $\mathcal{Q}_{max}$ in the MSSM run. In pink the points in the discovery region of directional detectors and in cyan in the exclusion region.}
\label{fig:MSSM_parameters_MIMAC}
\end{figure}

\begin{figure}[bt]
\centering
\includegraphics[width=0.5\textwidth ,natwidth=6cm,natheight=6cm]{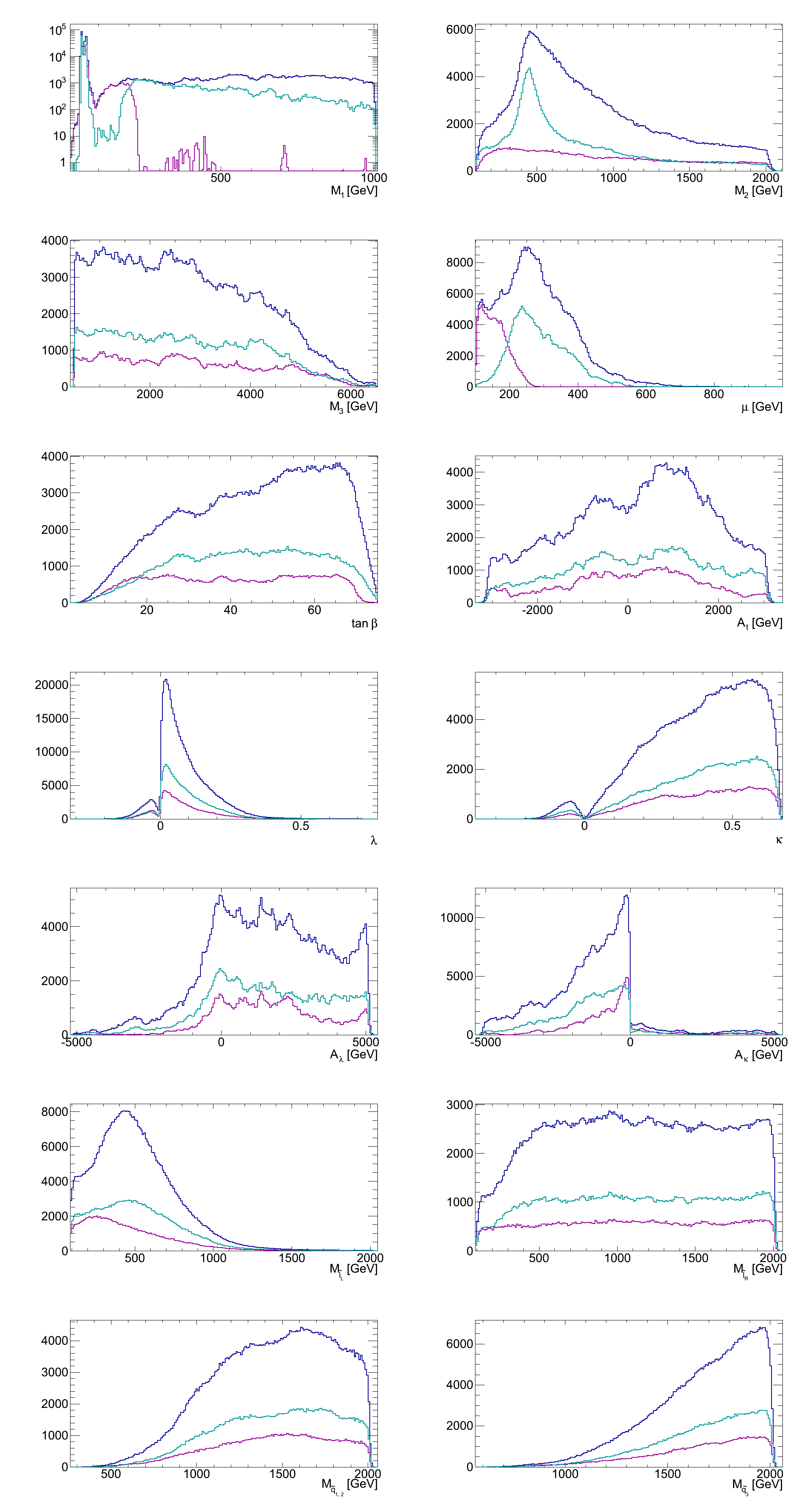}
\caption{Free parameter frequency distribution normalized to $\mathcal{Q}_{max}$ in the NMSSM run. In pink the points in the discovery region of directional detectors and in cyan in the exclusion region.}
\label{fig:NMSSM_parameters_MIMAC}
\end{figure}

Although the MSSM and NMSSM parameter spaces are multidimensional and can produce many different phenomenologies, the reach of directional detectors is determined solely by the DM particle mass and the SD cross section. In particular, for masses below 200~GeV, a large fraction of the supersymmetric configurations fall in the discovery region. For neutralinos to satisfy $m_{\chi_1^0}\leq200$ GeV, the mass term of at least one of its components has to satisfy the same condition. This can be either the bino component determined by $M_1$, the higgsino component given by $\mu$, the wino component given by $M_2$ or  the singlino component in the 
NMSSM (which depends on a combination of $\lambda$, $\kappa$ and $\mu$). Note however that when the neutralino LSP is wino-dominated, which implies $M_2 \leq M_1,\mu$, the relic density tends to be very small, those configurations are therefore excluded by the condition $\Omega_{\chi_1^0}h^2\geq10\%\Omega_{WMAP}h^2$. The neutralino LSP main component is therefore determined by the lighter of $M_1,\mu$. 
%
%Since we impose $\Omega_{\chi_1^0}h^2\geq10\%\Omega_{WMAP}h^2$, the two components that are allowed to %reach those values in the MSSM are the bino component determined by $M_1$ and the higgsino component given by $%\mu$, completed by the singlino component in the NMSSM (which depends on a combination of $\lambda$, $\kappa$% and $\mu$)\footnote{If one allows a negligible relic density, the neutralino LSP could be wino-dominated, which implies $M_2 \leq M_1,\mu$.}. The lighter of these two parameters represents the neutralino LSP main component. 
The strength of  the SD interactions, strongly depend on the $\mu$ value, determining the neutralino mixing elements $N_{13}$ and $N_{14}$, thus the Z$\chi_{1}^0\chi_{1}^0$ coupling. Since the lighter the $\mu$ the larger $\mid N_{13}\mid$ and $\mid N_{14}\mid$, the lighter the neutralino (although depending on $M_1$, and the singlino component in the NMSSM), we expect small $\mu$ values to be generally in the reach of directional detectors.

In Fig.~\ref{fig:mu_vs_m1_MIMAC} we present the correlation between $\mu$ and $M_1$, using the same color tag for points in or out the discovery and exclusion regions. The general trend just discussed is indeed met: a discovery is mostly linked with $\mu\lesssim$~250-300~GeV. Roughly speaking, the $\mu=M_1$ line separates the bino and higgsino dominated neutralino compositions.

For a bino-like neutralino to have large enough SD interactions (pink or cyan points), either the higgsino component is not negligible (i.e. points above the $\mu=M_1$ line but not far away from it), either the squarks have similar masses to those of neutralinos (which is the case of the few pink and cyan points above and far away from the $\mu=M_1$ line). The former case is very well represented by the vertical clouds at $M_1\sim$~50~GeV, corresponding to annihilation via the Z resonance. In such a line, $\mu$ below 300~GeV allows discovery (pink points), and up to $\sim$~450~GeV the configurations could be challenged by directional detectors (cyan points). Larger values of $\mu$ would be out of reach (blue points).

The effect of the neutralino mass is imprinted in the fact that very few pink points can be found at large values of $M_1$ and $\mu$. This is easily understood: the heavier the neutralino, the smaller the cross section and the smaller the projected sensitivity, the less configurations fall in the discovery region.

The correlation between small values of $\mu$ and the discovery potential is also observed in Figs.~\ref{fig:MSSM_parameters_MIMAC} and~\ref{fig:NMSSM_parameters_MIMAC}. In these figures we show the frequency distribution of all free parameters in the MSSM and NMSSM respectively for all points (blue), as well as for pink and cyan points only. In the fourth panel of both figures one can see that the pink distribution of $\mu$ is strongly peaked towards the lighter values, almost reaching the all point curve in the MSSM. 
Indeed, few configurations with $\mu\leq150$ GeV would actually escape a discovery with a large directional detector (30 kg.year). In the NMSSM such configurations could also have a large singlino component in which case they would remain out of reach of directional detectors. 
 Regarding $M_1$ (first panel), the pink and cyan distributions also reflect the neutralino mass range for detection and exclusion. In
 particular, apart from some configurations with $M_1$ around $M_Z/2$ (Z resonant neutralinos), most configurations with $M_1\lesssim450$
 GeV would be challenged by future directional detectors. Conversely, for $\mu,M_1\geq450$ GeV, most supersymmetric configuration would not be  probed by directional detectors, even if exclusion remains possible.
 %gb are you sure for NMSSM looks like there are a few. but hard to tell..

As for the other parameters, the shapes of distributions are to a large extent related to the LSP mass. 
The exclusion region does not contain any points with a LSP lighter than 35 GeV and extends to much larger masses for the LSP than the discovery region.  Thus  the cyan  distributions for the masses of supersymmetric 
particles will be shifted towards higher values than the pink distributions. In particular the distributions for the soft terms
$\mu,M_2$ that drive the masses of the neutralinos and charginos    are peaked at higher values for the exclusion region than for the discovery region.  For the same reason, 
the pink distribution for $M_{\tilde{l}_L}$ and $M_{\tilde{l}_R}$ are shifted towards smaller values than the cyan and blue distributions. 
Regarding $\tan\beta$,  we also expect  to see that smaller values are preferred for the points in the discovery region. Indeed, in the MSSM, the light neutralino LSP is often associated with a light $M_A$, collider constraints on the Higgs sector then selects small values of $\tan\beta$.
This trend is also met in the NMSSM, however, since the light neutralino is not necessarily associated with a light pseudoscalar mass, there is a much larger fraction of configurations with small values of $\tan\beta$ that escape SD detectability.
Finally we remark that the masses of squarks and gluinos can extend well above 1 TeV, this sector is mostly constrained by flavour observables. Therefore a large fraction of the configurations will escape detection at the LHC as will be discussed next.
%As for the other parameters, the shapes of distributions are related to the same phenomenon. Light neutralinos are often associated to light sleptons. Thus, the pink distribution for $M_{\tilde{l}_L}$ and $M_{\tilde{l}_R}$ are shifted towards smaller values than the cyan and blue distributions. In particular in the MSSM, the blue curve for $M_{\tilde{l}_L}$ can be interpreted as two peaks, one at 150-200 GeV and the main at 400 GeV. The pink distribution is peaked at the lighter, while the cyan distribution is peaked at the heavier. Regarding $\tan\beta$, it is also normal to see that smaller values are preferred for the points in the discovery region. Indeed, in the MSSM, the light neutralino LSP is often associated with a light $M_A$, collider constraints on the Higgs sector then selects small values of $\tan\beta$.
%This trend is also met in the NMSSM, however, since the light neutralino is not necessarily associated with a light pseudoscalar mass, there is a much larger fraction of configurations with small values of $\tan\beta$ that escapes SD detectability. 

\section{Impact of LHC results on the parameter space}\label{sec:LHC}
\label{sec:lhc}
\begin{figure}[htb]
\centering
\includegraphics[width=0.5\textwidth ,natwidth=6cm,natheight=6cm]{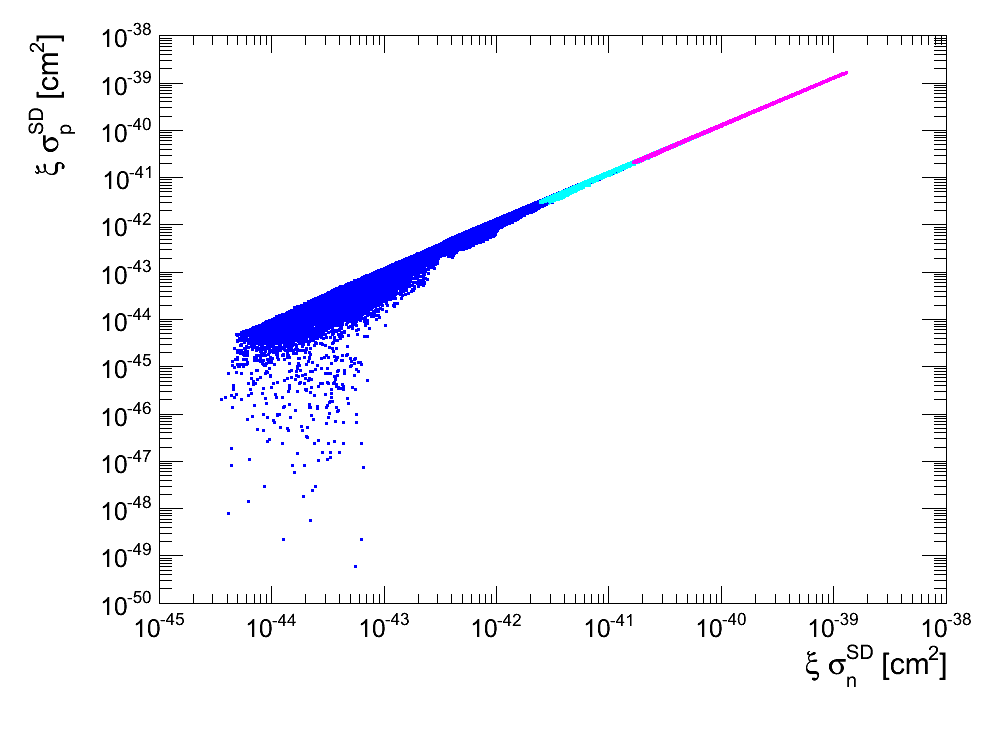}
\includegraphics[width=0.5\textwidth ,natwidth=6cm,natheight=6cm]{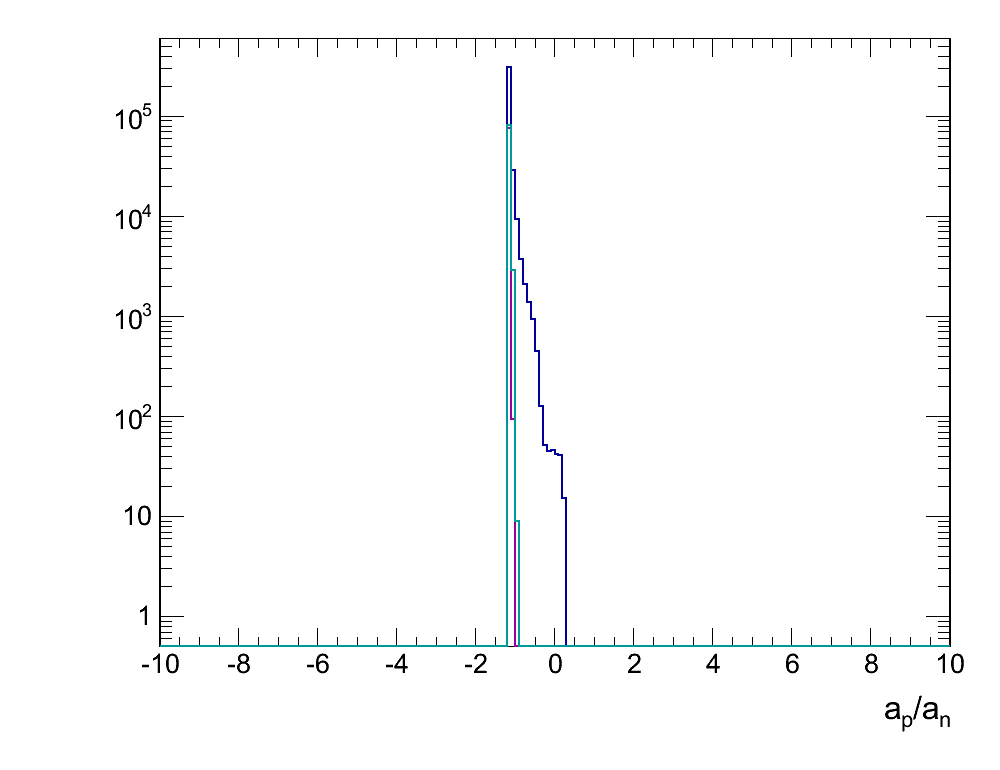}
\caption{Top Panel: Spin dependent elastic scattering cross sections correlations: 
proton-neutralino versus neutron-neutralino interactions in the MSSM. 
In pink the points in the discovery region of directional detectors and in cyan in the exclusion region. 
Here, in contrast with Figs.~\ref{fig:SDp_vs_SDn_MIMAC}, we have removed those points falling above the 
ATLAS limit on the $M_{\tilde{q}}$ vs. $M_{\tilde{g}}$ plane. Bottom panel: frequency distribution of $a_p/a_n$ for the same points with the same
color code.}
\label{fig:SDp_vs_SDn_MIMAC_safe}
\end{figure}

The ATLAS collaboration has published an analysis and established a limit on the squark mass versus gluino mass plane~\cite{Aad:2011ib}. 
These limits were derived for massless neutralino LSP in the MSSM and assuming that the squarks decay exclusively into a quark and the LSP. This is not the case in the scans we have performed in the MSSM, in particular additional decay modes of the squarks can weaken the ATLAS limits. However, comparing these limits to our data sets is a good way to check the possible influence of the forthcoming results of the LHC in terms of parameter space cutting and its implications for SD cross sections. The modification of the squark decay modes and the presence of light Higgs states mean that these limits do not apply to the NMSSM, thus we restrict this analysis to the 
MSSM\footnote{An analysis of the ATLAS constraints on the NMSSM parameter space is ongoing~\cite{NMSSM_ATLAS}.}.

We have applied the limit on the $M_{\tilde{q}}$ vs. $M_{\tilde{g}}$ plane, where $M_{\tilde{q}}$ stands for the lightest first or second
generation squark mass. Points falling below the curve provided by ATLAS are not represented in Fig.~\ref{fig:SDp_vs_SDn_MIMAC_safe} (top
panel), where we show the correlation between neutralino SD interactions with protons and neutrons. The difference is quite striking with respect to the top panel in Fig.~\ref{fig:SDp_vs_SDn_MIMAC}, where all points are drawn. What we observe is that, as the LHC probes the lightest squarks (in the hypothesis these are not observed), only the points in the main correlation line are left. Thus, the possibilities to achieve a large SD interaction would be restricted to 
a large higgsino fraction.\\ 
Bottom panel of Fig.~\ref{fig:SDp_vs_SDn_MIMAC_safe} presents the frequency distribution of $a_p/a_n$ for the same points with the same
color code. As outlined in sec.~\ref{sec:SDCross}, the relative sign of $a_p$ and $a_n$ is a 
key issue  for SD direct searches. It is worth noticing that taking into account recent ATLAS limits on squarks and gluinos 
restricts the allowed models to the ones with 
negative values of $a_p/a_n$. More precisely, only the Z exchange is allowed, and the ratio of the two amplitudes is given by  
$a_p/a_n \simeq-1.14$ (see sec.~\ref{sec:xs1}).  It implies that a nucleus target with spin contents of opposite 
sign ({\it e.g.} $\rm ^{19}F$, $\rm ^{3}He$ or $\rm ^{133}Cs$) will present a constructive interference while nucleus target with spin contents 
of same sign ({\it e.g.} $\rm ^{129}Xe$, $\rm ^{131}Xe$ or $\rm ^{73}Ge$) will have a destructive interference thus reducing the cross
section on nucleus and hence the event rate. A dedicated study is needed on this issue, but we emphasize that it may have consequences in
the choice of target for upcoming SD experiments.
Note that this statement is rather independent of the details of the SUSY model, as long as LHC exclusions on squarks apply. 

%fm inutile ? 
%Also, the smallest allowed nucleon cross section would be $\geq 10^{-45}$ cm$^2$. This is, however, still far below the reach of 
%current and forthcoming detectors.

\section{A discovery scenario}

\begin{figure}[hbt]
\centering
\includegraphics[width=0.5\textwidth ,natwidth=6cm,natheight=6cm]{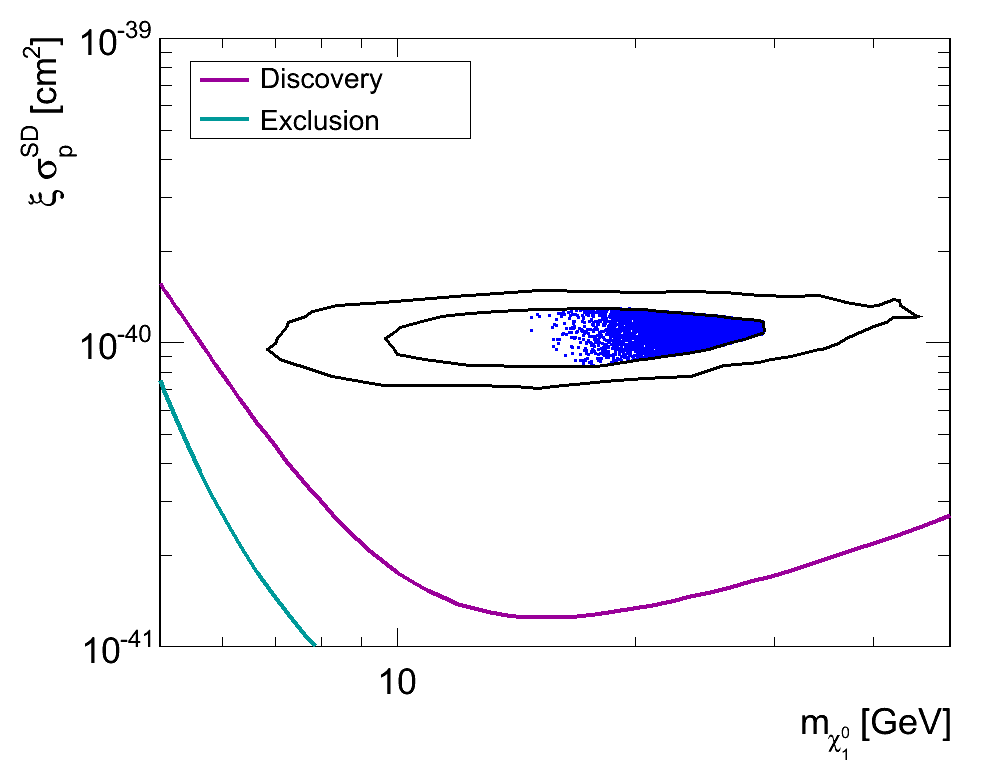}
\includegraphics[width=0.5\textwidth ,natwidth=6cm,natheight=6cm]{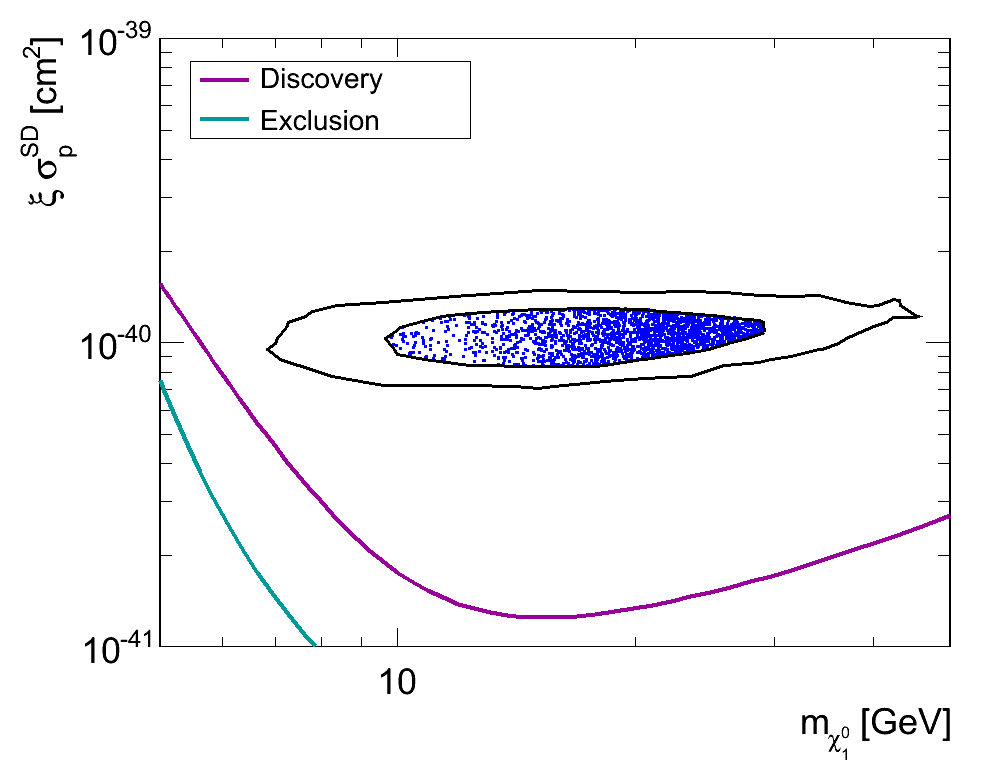}
\caption{Proton-neutralino spin dependent elastic scattering cross section versus the neutralino mass in the discovery scenario runs. Top: MSSM. Bottom: NMSSM. We display the 1$\sigma$ and 3$\sigma$ contours, while we used the former as the constraint for the random walk. Also displayed are the exclusion and discovery projections for a nominal directional detector. We display only safe points regarding XENON100, Fermi-LAT and CMS (MSSM only) limits on SI elastic scattering interactions, $\gamma$-rays from the Draco dSph, and Higgs interactions.}
\label{fig:SDp_contour}
\end{figure}

\label{sec:disco}
\begin{figure}[hbt]
\centering
\includegraphics[width=0.44\textwidth ,natwidth=6cm,natheight=5.cm]{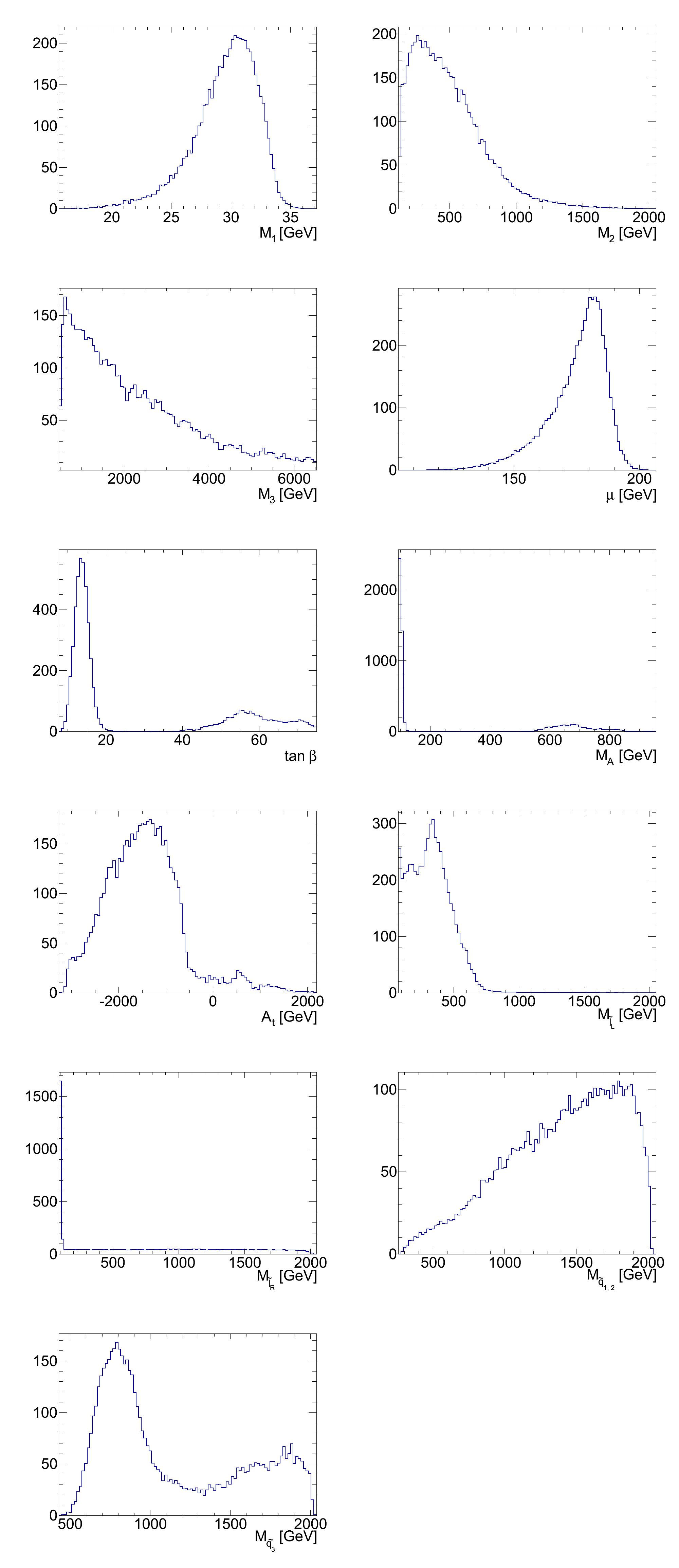}
\caption{Free parameter frequency distribution normalized to $\mathcal{Q}_{max}$ in the MSSM discovery scenario run.}
\label{fig:MSSM_parameters}
\end{figure}

\begin{figure}[hbt]
\centering
\includegraphics[width=0.5\textwidth ,natwidth=6cm,natheight=6cm]{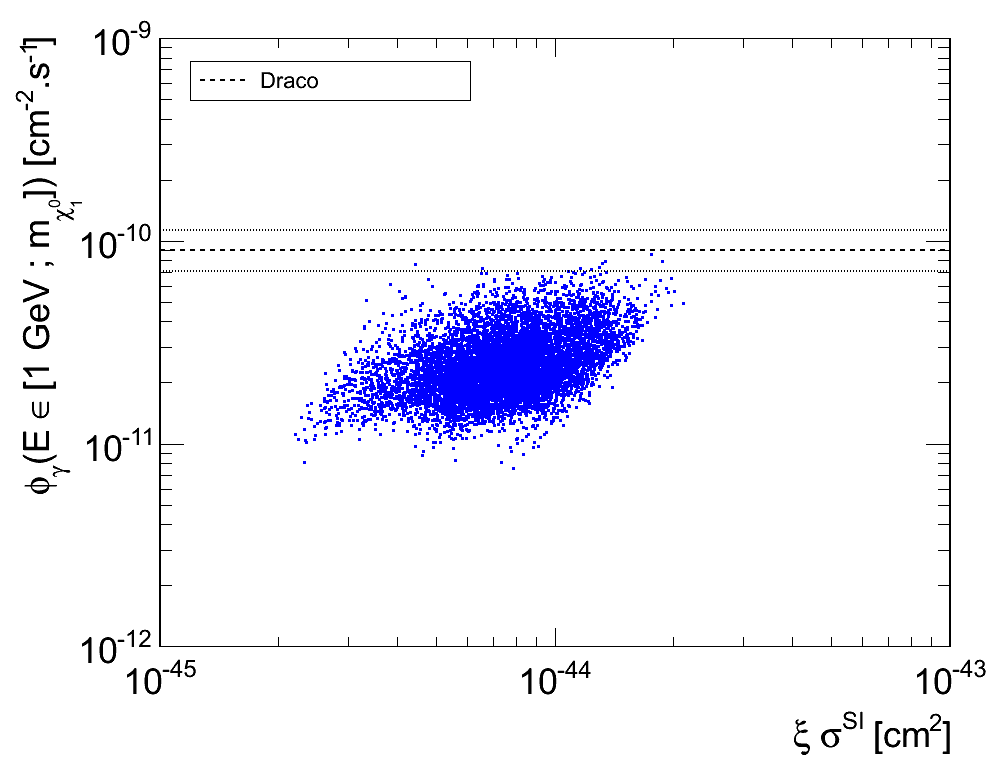}
\includegraphics[width=0.5\textwidth ,natwidth=6cm,natheight=6cm]{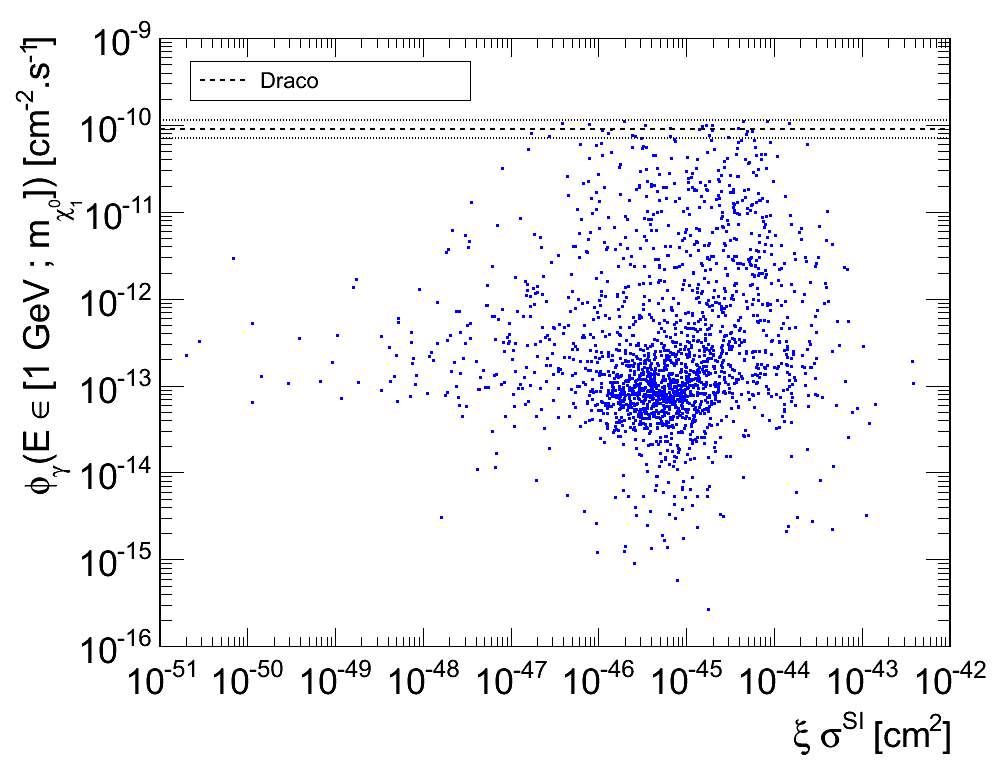}
\caption{Flux of $\gamma$-rays expected from neutralino annihilations from the Draco dwarf spheroidal galaxy versus spin independent cross section. Top: MSSM. Bottom: NMSSM. The Fermi-LAT limits~\cite{Abdo:2010ex} are also displayed. Only the points overcoming all constraints (from XENON100, Fermi-LAT and CMS) in the discovery scenario are displayed.}
\label{fig:Dwarf_SI_contour}
\end{figure}

As outlined in sec.~\ref{sec:DMprop}, a dedicated analysis of data of a 
30 kg.year $\rm CF_4$ directional detector could also allow us to constrain  
the  WIMP properties, both from particle physics ($m_\chi, \sigma_p^{SD}$) and 
galactic halo (velocity dispersions)~\cite{billard.ident}. This would be one step beyond current DM search strategy capabilities. Of
course, this requires a rather large SD cross section associated with a low neutralino mass. The outcome would be a constraint on the mass
and the cross section, {\it i.e.} within a contour defined by a confidence level.\\ 
%If a directional detector observes DM events, then it is possible to give a range of DM interaction rates and masses, 
%as we discussed in Sec.~\ref{sec:DMprop}. In particular, 
%a canonical directional detector would resolve the mass and cross section system within a contour defined by a confidence level. 
%This contour is what a directional detection experiment could yield as an experimental result.
 In light of such a tool, we perform the following exercise~\cite{billard.ident}: we assume the existence of a DM particle of 20 GeV mass with $10^{-4}$ pb interaction 
 rate with protons, leading to $\approx 80$ WIMP events in a 30 kg.year $CF_4$ directional detector, to generate simulated data. Then, we infer the contour from the data analysis procedure after detection by a canonical directional detector. We exploit this result by including a new prior in the MCMC described in~\ref{sec:Meth}: we impose the neutralino mass and the SD cross section to lie in the 1$\sigma$ contour obtained by the discovery. Hence, we are able to scan the possible scenarios fitting the observation.

%We carried out this application for both the MSSM and the NMSSM, with the same set of free parameters and applying the same constraints 
%as   before. We fixed the starting points of the random walks by picking among previous points the ones lying within the contour. This induces a %statistical bias, however, we chose the initial points such that they were as different from each other as possible, and let 10 chains evolve per %model, each with a different starting point.

In Fig.~\ref{fig:SDp_contour} we show the distribution of points within the contour in the SD vs. neutralino mass plane. Notice that the transition in the contour prior was sharp, so no point is kept outside those boundaries. In the NMSSM (bottom panel), we have much less statistics than in the MSSM (top panel), however, populating more this discovery region is only a matter of time. Nevertheless, the contour is much more homogeneously filled in the NMSSM case, showing that there are more possibilities to find good configurations, in contrast with the MSSM were the heavy end of the contour is much more preferred than the actual value of the WIMP in the simulation input. Hence, the NMSSM neutralino DM appears a much more plausible explanation if such a detection was made. In these figures we have kept only the points that satisfied all constraints including the CMS limits on $\tan\beta$ vs. $M_A$ plane (MSSM only) as well as  Fermi-LAT  and XENON100 limits.

An observation would allow to refine the expectations on the soft SUSY parameters, as can be seen by comparing the parameter distribution in Fig.~\ref{fig:MSSM_parameters_MIMAC} with the one corresponding to points in the contour in Fig.~\ref{fig:MSSM_parameters} in the  case of the MSSM. 
Most distributions are determined by the condition on the  neutralino LSP mass. This is clearly the case for $M_1$ which drives the mass of the  light bino,  but also applies to other parameters~\cite{Vasquez:2011yq}. Indeed for a light LSP to be consistent with 
the relic density constraint  requires either a light slepton, hence the very peaked distribution in for $M_{\tilde{l}_R}$ or
a light Higgs boson, hence the peak in the $M_A$ distribution. Higgs and flavor physics constraints then imply that
intermediate values of $\tan\beta$ are disfavored. Furthermore the squark contribution that is needed to cancel the Higgs contribution in $B(b\rightarrow s\gamma)$ explains the peak at low values of the third generation squark masses $M_{\tilde{q}_3}$. The main impact of imposing a specific range for $\sigma^{SD}$ is reflected in the distribution for $\mu$, confined to be in a narrow range.  As mentioned previously, the strength of the neutralino coupling to the Z is the most relevant parameter in computing $\sigma^{SD}$. 

%Indeed, $M_1$ is given by the neutralino mass, which has to be mostly bino. As we had already commented, one way %of achieving good MSSM configurations in the discovery mass range is by having the slepton mass has to be as light as possible, hence the very peaked distribution in for $M_{\tilde{l}_R}$, and the other is light Higgs bosons, hence the peak in the $M_A$ distribution. These distributions could be use to study different collider predictions.

Figures~\ref{fig:Dwarf_SI_contour} represent the safe points  in the $\gamma$-rays from Draco 
vs. SI interaction plane. There is a concentration of points at $10^{-46}-10^{-45}$ cm$^2$ cross sections and $10^{-13}\; \rm cm^3\;s^{-1}$ fluxes. This constitutes the prediction for other detection techniques in case an observation is made. Due to the interplay of parameters, the spread of these predictions is rather large in the NMSSM: 9 orders of magnitude in the SI axis and 6 orders of magnitude in $\phi_\gamma$. An order of magnitude spread is predicted in the MSSM. This is an extremely different behavior of the two models. Such a discovery would 
predict a MSSM neutralino to be found shortly by direct detection and indirect detection experiments, thus provinding a cross check of the
discovery claim. 
On the contrary, a NMSSM neutralino could easily escape any other detection.

In any case, keep in mind that such a discovery would still mean that $\mu \lesssim 200$ GeV, and $10\;\rm{GeV} \lesssim m_{\chi_1^0}\lesssim 30\;\rm{GeV}$. With these two characteristics we can predict the presence of a chargino with $m_{\chi_1^+}\lesssim 200$ GeV. A light slepton would be favored in the MSSM case, but its absence leaves only the NMSSM as a possibility. With the help of collider physics, it could be possible to rule out the possibility of a MSSM neutralino DM, or even of any supersymmetric neutralino if no charged particle is observed in the vicinity of the weak scale.

\section{Conclusion}
We have shown that supersymmetric models with neutralino DM predict signals in the range of detection for SI and SD direct detectors  as well as for indirect detectors. 
However only a fraction of the parameter space of either the MSSM or the NMSSM can be probe by each type of experiments alone. The development of SD directional detection techniques thus offer the possibility to probe more thoroughly the parameter space of supersymmetric models. 
This is the case even if the sensitivity of indirect detectors as well as SI detectors is increased by one order (or more)  of magnitude, thus emphasizing the complementarity between different techniques. 

With the planned MIMAC detector,  neutralinos up to 200 GeV could be discovered and up to 600 GeV could be excluded. 
The light neutralinos  that are best probed by directional detectors are often accompanied by not so heavy  charginos/neutralinos and even sleptons or by a light pseudoscalar. 
The search for these weakly interacting particles at the LHC will therefore impact  in the future the potential of directional detection to probe supersymmetric models.  
We expect the LHC to considerably expand the constraints on supersymmetric scenarios in the next year, even if no signal of physics beyond the standard model is found. We have shown that if squarks of the first two generations are excluded up to a mass of nearly 750~GeV, notwithstanding that they could escape detection due to small mass splittings with some other supersymmetric particles, the squarks play little role in direct detection and the SD cross section on protons is completely correlated with that on neutrons. 
The LHC is also probing supersymmetric models with Higgs searches, in particular the negative search results on the 
heavy Higgs doublet of the MSSM  constrain the supersymmetric models with a light neutralino. 
Furthermore a confirmation of a Higgs  signal at 125 GeV as announced by ATLAS would narrow down the number of allowed supersymmetric configurations. Note however that the light Higgs is not directly linked with SD direct detection. 

\section*{Acknowledgment}
GB thanks the LPSC where part of this work was done, for its hospitality. DAV thanks C\'eline B\oe hm for very useful discussions and advice.\\

\end{document}